\documentclass[journal]{vgtc}                

\ifpdf
  \pdfoutput=1\relax                   
  \usepackage{graphicx}                
  \DeclareGraphicsExtensions{.pdf,.png,.jpg,.jpeg} 
\else
  \usepackage{graphicx}                
  \DeclareGraphicsExtensions{.eps}     
\fi%

\usepackage[acronym,shortcuts]{glossaries}

\graphicspath{{figures/}{pictures/}{images/}{./}} 

\usepackage{microtype}                 
\PassOptionsToPackage{warn}{textcomp}  
\usepackage{textcomp}                  
\usepackage{mathptmx}                  
\usepackage{times}                     
\usepackage{cite}                      
\usepackage{tabu}                      
\usepackage{booktabs}                  
\usepackage[skins]{tcolorbox}
\usepackage{siunitx}
\usepackage{enumitem}
\usepackage[normalem]{ulem}
\usepackage{amssymb}
\usepackage{ccicons}

\usepackage[hang,scriptsize,sf]{subfigure}
\onlineid{0}

\vgtccategory{Research}
\vgtcpapertype{algorithm/technique}

\title{ScaleTrotter: Illustrative Visual Travels Across Negative Scales}

\author{Sarkis Halladjian, Haichao Miao, David Kou\v{r}il, M. Eduard Gr{\"o}ller, Ivan Viola, and Tobias Isenberg}
\authorfooter{
\item Sarkis Halladjian and Tobias Isenberg are with Inria, France. E-mail: \{sarkis.halladjian\,$|$\,tobias.isenberg\}@inria.fr.
\item Sarkis Halladjian is also with Universit{\'e} Paris-Saclay, France.
\item Haichao Miao, David Kou\v{r}il, and M. Eduard Gr{\"o}ller are with TU Wien, Austria. E-mail: \{miao\,$|$\,dvdkouril\,$|$\,groeller\}@cg.tuwien.ac.at.
\item Eduard Gr{\"o}ller is also with the the VRVis Research Center, Austria
\item Ivan Viola is with KAUST: King Abdullah University of Science and Technology, Kingdom of Saudi Arabia. E-mail: ivan.viola@kaust.edu.sa.
}

\shortauthortitle{Halladjian \MakeLowercase{\textit{et al.}}: ScaleTrotter---Multi-Scale Visualization of the Entire Human Genome}

\abstract{We present ScaleTrotter, a conceptual framework for an interactive, multi-scale visualization of biological mesoscale data and, specifically, genome data. ScaleTrotter allows viewers to smoothly transition from the nucleus of a cell to the atomistic composition of the DNA, while bridging several orders of magnitude in scale. The challenges in creating an interactive visualization of genome data are fundamentally different in several ways from those in other domains like astronomy that require a multi-scale representation as well. First, genome data has intertwined scale levels---the DNA is an extremely long, connected molecule that manifests itself at all scale levels. Second, elements of the DNA do not disappear as one zooms out---instead the scale levels at which they are observed group these elements differently. Third, we have detailed information and thus geometry for the entire dataset and for all scale levels, posing a challenge for interactive visual exploration. Finally, the conceptual scale levels for genome data are close in scale space, requiring us to find ways to visually embed a smaller scale into a coarser one. We address these challenges by creating a new multi-scale visualization concept. We use a scale-dependent camera model that controls the visual embedding of the scales into their respective parents, the rendering of a subset of the scale hierarchy, and the location, size, and scope of the view. In traversing the scales, ScaleTrotter is roaming between 2D and 3D visual representations that are depicted in integrated visuals. We discuss, specifically, how this form of multi-scale visualization follows from the specific characteristics of the genome data and describe its implementation. Finally, we discuss the implications of our work to the general illustrative depiction of multi-scale data.}

\CCScatlist{ 
 \CCScat{K.6.1}{Management of Computing and Information Systems}%
{Project and People Management}{Life Cycle};
 \CCScat{K.7.m}{The Computing Profession}{Miscellaneous}{Ethics}
}

\teaser{
	\centering
	\includegraphics[width=\linewidth]{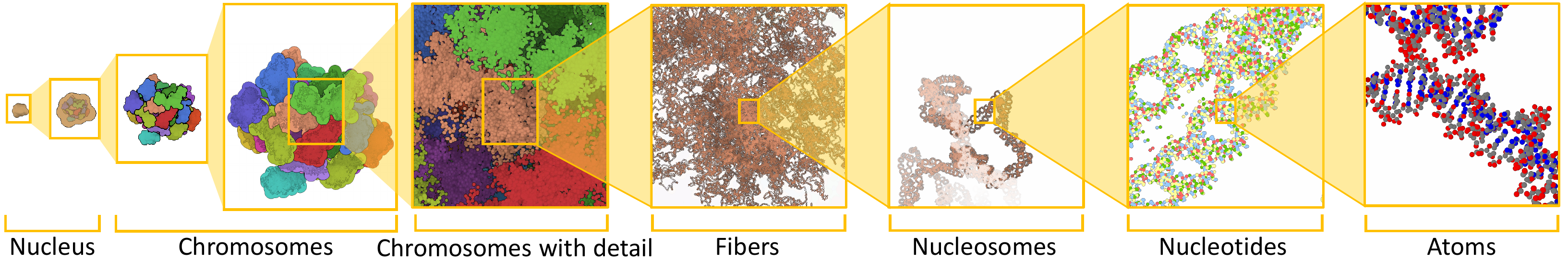}\vspace{-1ex}
	\caption{Steps along the voyage into the genomic detail as enabled by ScaleTrotter, showing the semantic scale levels. \protect\new{ScaleTrotter and its \emph{visual embedding} allow us to seamlessly transition between independent representations and interactively explore them.}}
	\label{fig:teaser}
}

\definecolor{myblue1}{rgb}{0.1, 0.6, 0.8}
\definecolor{myred1}{rgb}{0.7, 0.0, 0.0}
\definecolor{mygreen1}{rgb}{0.0, 0.8, 0.0}
\definecolor{mypurple1}{rgb}{0.6, 0.0, 0.6}
\definecolor{mycolor1}{rgb}{0.0, 0.6, 0.6}

\newcommand{\new}[1] {\textcolor{red}{#1}} 
\renewcommand{\new}[1]{#1}

\newcommand{\eg}{e.\,g.}
\newcommand{\ie}{i.\,e.}
\newcommand{\etal}{~et~al.}
\newcommand{\emphbf}[1]{\emph{\textbf{#1}}}

\setacronymstyle{long-short}
\newacronym{LOD}{LOD}{level-of-detail}

\vgtcinsertpkg

\begin{document}


\firstsection{Introduction}

\maketitle

\label{sec:intro}

The recent advances in visualization have allowed us to depict and understand many aspects of the structure and composition of the living cell. For example, cell\-VIEW \cite{LeMuzic:2015:CEL} provides detailed visuals for viewers to understand the composition of a cell in an interactive exploration tool and Lindow et al. \cite{Lindow:2019:IVR} created an impressive interactive illustrative depiction of RNA and DNA structures. Most such visualizations only provide a depiction of components/processes at a single scale level. Living cells, however, comprise structures that function at scales that range from the very small to the very large. The best example is DNA, which is divided and packed into visible chromosomes during mitosis and meiosis, while being read out at the scale level of base pairs. 
In between these scale levels, the DNA's structures are typically only known to structural biologists, while beyond the base pairs their atomic composition has implications for specific DNA properties.

The amount of information stored in the DNA is enormous. The human genome consists of roughly 3.2\,Gb (giga base pairs) \cite{Alberts:2015:MBC6,Schneider:2017:EG}. This information would fill 539,265 pages of the TVCG template,
which would stack up to approx.\ \SI{27}{\metre}.
Yet, the whole information is contained inside the cell's nucleus with only approx.\ \SI{6}{\micro\metre} diameter \cite[page 179]{Alberts:2015:MBC6}. 
\new{Similar to a coiled telephone cord, the DNA creates a compact structure that contains the long strand of genetic information. This organization results in several levels of perceivable structures (as shown in \autoref{fig:teaser}), which have been studied and visualized separately in the past.}
The problem thus arises of how to comprehend and explore \new{the whole scope of this} massive amount of multi-scale information. 
If we teach students or the general public about the relationships between the two extremes, for instance, we have to ensure that they understand how the different scales work together. Domain experts, in contrast, deal with questions such as whether correlations exist between the spatial vicinity of bases and genetic disorders. It may manifest itself through two genetically different characteristics that are far from each other in sequence but close to each other in the DNA's 3D configuration. For experts we thus want to ensure that they can access the information at any of the scales. They should also be able to smoothly navigate the information space. 
\new{The fundamental problem is thus to understand how we can enable a smooth and intuitive navigation in space and scale with seamless transitions. For this purpose we derive specific requirements of multi-scale domains and data with negative scale exponents and analyze how the constraints affect their representations.}


\new{Based on our analysis we} introduce \new{ScaleTrotter,} an interactive multi-scale visualization of the human DNA, ranging from the level of the interphase chromosomes\footnote{While interphase chromosomes were measured to be approx.\ \SI{12}{\micro\metre} long, this was after ``flattening'' and ``routine chromosome preparation'' \cite{Lemke:2002:DSH}.} in the \SI{6}{\micro\metre} nucleus to the level of base pairs ($\approx$\,\SI{2}{\nano\metre}) resp.\ atoms ($\approx$\,\SI{0.12}{\nano\metre}). We cover a scale range of 4--5 orders of magnitude in spatial size, and allow viewers to interactively explore as well as smoothly interpolate between the scales. We focus specifically on the visual transition between neighboring scales, so that viewers can mentally connect them and, ultimately, understand how the DNA is constructed. 
With our work we go beyond existing multi-scale visualizations due to the DNA's specific character. Unlike multi-scale data from other fields, the DNA physically connects conceptual elements across all the scales \new{(like the phone cord)} so it never disappears from view. We also need to show detailed data everywhere and, for all stages, the scales are close together in scale space.


We base our implementation on multi-scale data from genome research about the positions of DNA building blocks, which are given at a variety of different scales. We then transition between these levels using what we call visual embedding. It maintains the context of larger-scale elements while adding details from the next-lower scale. We combine this process with scale-dependent rendering that only shows relevant amounts of data on the screen. Finally, we support interactive data exploration through scale-dependent view manipulations, interactive focus specification, and visual highlighting of the zoom focus.




In summary, our contributions are as follows.
First, we analyze the unique requirements of multi-scale representations of genome data and show that they cannot be met with existing approaches.
Second, we demonstrate how to achieve smooth scale transitions for genome data through visual embedding of one scale within another based on measured and simulated data. We further limit the massive data size with a scale-dependent camera model to avoid visual clutter and to facilitate interactive exploration.
Third, we describe the implementation of this approach and compare our results to existing illustrations\new{. Finally, we} report on feedback from professional illustrators \new{and domain experts. It indicates that our interactive visualization can serve as a fundamental building block for tools that target both domain experts and laypeople.}\vspace{-.25pt}

\section{Related Work}

Our work concerns the use of abstraction in illustrative visualization, visual representations of small-scale biology and genome data, and general multi-scale data visualization techniques, as we discuss next.\vspace{-.25pt}

\subsection{Abstraction in illustrative visualization}

On a high level, our work relates to the use of abstraction in creating effective visual representations, \ie, the use of \emph{visual abstraction}. Viola and Isenberg \cite{Viola:2018:PCA} describe this concept as a process,  which removes detail when transitioning from a lower-level to a higher-level representation, yet which preserves the overall concept. While they attribute the removed detail to ``natural variation, noise, etc.'' in the investigated multi-scale representation we actually deal with a different data scenario: DNA assemblies at different levels of scale. We thus technically do not deal with a ``concept-preserving transformation'' \cite{Viola:2018:PCA}, but with a process in which the underlying representational concept (or parts of it) can change. Nonetheless, their view of abstraction as an interactive process that allows viewers to relate one representation (at one scale) to another one (at a different scale) is essential to our work.

Also important from Viola and Isenberg's discussion \cite{Viola:2018:PCA} is their concept of \emph{axes of abstraction}, which are traversed in scale space. We also connect the DNA representations at different scales, facilitating a smooth transition between them. In creating this axis of abstraction, we focus primarily on changes of Viola and Isenberg's geometric axis, but without a geometric interpolation of different representations. Instead, we use visual embedding of one scale in another one.

\subsection{Scale-dependent molecular and genome visualization}
\label{sec:rw:biomolecular}

We investigate multi-scale representations of the DNA, which relates to work in bio-molecular visualization. Several surveys have summarized work in this field \cite{Kozlikova:2015:VBS, Alharbi:2017:VCC,Kozlikova:2017:VBS,Miao:2019:MMV}, so below we only point out selected approaches. In addition, a large body of work by professional illustrators on mesoscale cell depiction inspired us such as visualizing the human chromosome down to the detail of individual parts of the molecule \cite{Goodsell:2018:FAC}.


In general, as one navigates through large-scale 3D scenes, the underlying subject matter is intrinsically complex and requires appropriate interaction to aid intellection \cite{Glueck:2011:CMS}. The inspection of individual parts is challenging, in particular if the viewer is too far away to appreciate its visual details. Yet large, detailed datasets or procedural approaches are essential to create believable representations. To generate not only efficient but \emph{effective} visualizations, we thus need to remove detail in Viola and Isenberg's \cite{Viola:2018:PCA} visual abstraction sense. This allows us to render at interactive rates as well as to see the intended structures, which would otherwise be hidden due to cluttered views. Consequently, even most single-scale small-scale representations use some type of multi-scale approach and with it introduce abstraction. Generally we can distinguish three fundamental techniques: multi-scale representations by leaving out detail of a single data source, multi-scale techniques that actively represent preserved features at different scales, and multi-scale approaches that can also transit between representations of different scales. We discuss approaches for these three categories next.

\subsubsection{Multi-scale visualization by means of leaving out detail}

An example of leaving out details in a multi-scale context is Parulek et al.'s \cite{Parulek:2014:CLD} continuous levels-of-detail for large molecules and, in particular, proteins. They reduced detail of far-away structures for faster rendering. They used three different conceptual distances to create increasingly coarser depictions such as those used in traditional molecular illustration. For distant parts of a molecule, in particular, they seamlessly transition to super atoms using implicit surface blending.

The cell\-VIEW framework \cite{LeMuzic:2015:CEL} also employs a similar \ac{LOD} principle using advanced GPU methods for proteins in the HIV. It also removes detail to depict internal structures, and procedurally generates the needed elements.
In mesoscopic visualization, Lindow\etal \cite{Lindow:2012:IRM} applied grid-based volume rendering to sphere raycasting to show large numbers of atoms. They bridged five orders of magnitude in length scale by exploiting the reoccurrence of molecular sub-entities. 
Finally, Falk\etal \cite{Falk:2013:AVM} proposed out-of-core optimizations for visualizing large-scale whole-cell simulations. Their approach extended Lindow\etal's \cite{Lindow:2012:IRM} work and provides a GPU ray marching for triangle rendering to depict pre-computed molecular surfaces. 



Approaches in this category thus create a ``glimpse'' of multi-scale representations by removing detail and adjusting the remaining elements accordingly. We use this principle, in fact, in an extreme form to handle the multi-scale character of the chromosome data. We completely remove the detail of a large part of the dataset. If we would show all small details, an interactive rendering would be impossible and they would distract from the depicted elements. Nonetheless, this approach typically only uses a single level of data and does not incorporate different conceptual levels of scale.

\subsubsection{Different shape representations by conceptual scale}

The encoding of structures through different \emph{conceptual scales} is often essential.
Lindow\etal \cite{Lindow:2019:IVR}, for instance, described different rendering methods of nucleic acids---from 3D tertiary structures to linear 2D and graph models---with a focus on visual quality and performance. They demonstrate how the same data can be used to create both 3D-spatial representations and abstract 2D mappings of genome data. This produces three scale levels: the actual sequence, the helical form in 3D, and the spatial assembly of this form together with proteins.
Waltemate\etal \cite{Waltemate:2014:MMC} represented the mesoscopic level with meshes or microscopic images, while showing detail through molecule assemblies. To transition between the mesoscopic and the molecular level, they used a membrane mapping to allow users to inspect and resolve areas on demand. A magnifier tool overlays the high-scale background with lower-scale details. This approach relates to our transition scheme, as we depict the higher scale as background and the lower scale as foreground. 
A texture-based molecule rendering has been proposed by Bajaj\etal \cite{Bajaj:2004:Texmol}. Their method reduces the visual clutter at higher levels by incorporating a biochemically sensitive \ac{LOD} hierarchy.

Tools used by domain experts also visualize different conceptual genome scales. To the best of our knowledge, the first tool to visualize the 3D human genome has been Genome3D \cite{Asbury:2010:G3D}. It allows researchers to select a discrete scale level and then load data specifically for this level. The more recent GMOL tool \cite{Nowotny:2015:GMO} shows 3D genome data captured from Hi-C data \cite{vanBerkum:2010:HMS}. GMOL uses a six-scale system similar to the one that we employ and we derived our data from theirs. They only support a discrete ``toggling between scales'' \cite{Nowotny:2015:GMO}, while we provide a smooth scale transition. Moreover, we add further semantic scale levels at the lower end to connect base locations and their atomistic compositions.

\subsubsection{Conceptual scale representations with smooth transition}

A smooth transition between scales has previously been recognized as important. For instance, van der Zwan et al.\ \cite{Zwan:2011:IMV} carried out structural abstraction with seamless transitions for molecules by continuously adjusting the 3D geometry of the data. Miao\etal \cite{Miao:2018:MVS} substantially extended this concept and applied it to DNA nanostructure visualization. They used ten semantic scales and defined smooth transitions between them. This process allows scientists to interact at the appropriate scale level. Later, Miao et al.\ \cite{Miao:2018:DDS} combined this approach with three dimensional embeddings. In addition to temporal changes of scale, Lueks et al. \cite{Lueks:2011:SCC} explored a seamless and continuous \emph{spatial} multi-scale transition by geometry adjustment, controlled by the location in image or in object space.
Finally, Kerpedjiev\etal \cite{Kerpedjiev:2018:HWV} demonstrated multi-scale navigation of 2D genome maps and 1D genome tracks employing a smooth transition for the user to zoom into views.

All these approaches only transition between nearby scale levels and manipulate the depicted data geometry, which limits applicability. These methods, however, do not work in domains where a geometry transition cannot be defined. Further, they are limited in domains where massive multi-scale transitions are needed due to the large amount of geometry that is required for the detailed scale levels. We face these issues in our work and resolve them using visual embeddings instead of geometry transitions as well as a scale-dependent camera concept. Before detailing our approach, however, we first discuss general multi-scale visualization techniques from other visualization domains.\vspace{-.5pt}



\subsection{General multi-scale data visualization}

The vast differences in spatial scale of our world in general have fascinated people for a long time. Illustrators have created explanations of these scale differences in the form of images (\eg, \cite{Watzke:2017:MSU} and \cite[Fig.~1]{Pennisi:2001:HG}), videos (\eg, \href{http://www.eamesoffice.com/the-work/powers-of-ten/}{the seminal ``Powers of Ten'' video} \cite{Eameses:1977:PTR} from 1977), and newer \href{https://learn.genetics.utah.edu/content/cells/scale/}{interactive experiences} (\eg, \cite{GSLC:CSS}). Most illustrators use a smart composition of images blended such that the changes are (almost) unnoticeable, while some use clever perspectives to portray the differences in scale. These inspirations have prompted researchers in visualization to create similar multi-scale experiences, based on real datasets.

The classification from \autoref{sec:rw:biomolecular} for molecular and genome visualization applies here as well. Everts et al.\ \cite{Everts:2015:EBW}, \eg, \emph{removed detail} from brain fiber tracts to observe the characteristics of the data at a higher scale. Hsu et al.\ \cite{Hsu:2011:RFM} defined various cameras for a dataset, each showing a different level of detail. They then used image masks and camera ray interpolation to create smooth spatial scale transitions that show the data's multi-scale character.
Next, Glueck et al.\ \cite{Glueck:2009:MRV}'s approach exemplifies the \emph{change of shape representations by conceptual scale} by smoothly changing a multi-scale coordinate grid and position pegs to aid depth perception and multi-scale navigation of 3D scenes. They simply remove detail for scales that no longer contribute much to the visualization. In \href{https://www.youtube.com/watch?v=iMp_3c8OKgA}{their accompanying video}, interestingly, they limited the detail for each scale to only the focus point of the scale transition to maintain interactive frame rates. Another example of this category are geographic multi-scale representations such as online maps (\eg, Google or Bing maps), which contain multiple scale representations, but typically toggle between them as the user zooms in or out.
Finally, virtual globes are an example for \emph{conceptual scale representations with smooth transitions}. They use smooth texture transitions to show an increasing level of detail as one zooms in. Another example is Mohammed et al.'s \cite{Mohammed:2018:AVT} Abstractocyte tool, which depicts differently abstracted astrocytes and neurons. It allows users \href{https://www.youtube.com/watch?v=ARjQ_W97KKI}{to smoothly transition between the cell-type abstractions using both geometry transformations and blending}. We extend the latter to our visual embedding transition.

Also these approaches only cover a relatively small scale range. Even online map services cover less than approx.\ six orders of magnitude. Besides the field of bio-molecular and chemistry research discussed in \autoref{sec:rw:biomolecular}, in fact, only astronomy deals with large scale differences. Here, structures range from celestial bodies (\mbox{$\geq \: \approx 10^2$\,m})\footnotemark\ to the size of the observable universe ($1.3\cdot10^{26}$\,m), in total 24 orders of magnitude.

To depict such data, visualization researchers have created explicit multi-scale rendering architectures. Schatz et al.\ \cite{Schatz:2016:IVE}, for example, combined the rendering of overview representations of larger structures with the detailed depiction of parts that are close to the camera or have high importance. To truly traverse the large range of scales of the universe, however, several datasets that cover different orders of size and detail magnitude have to be combined into a dedicated data rendering and exploration framework.
The first such framework was introduced by Fu et al.\ \cite{Hanson:2000:VLS,Fu:2007:TSV} who used scale-independent modeling and rendering and power-scaled coordinates to produce scale-insensitive visualizations. This approach essentially treats, models, and visualizes each scale separately and then blends scales in and out as they appear or disappear. The different scales of entities in the universe can also be modeled using a \emph{ScaleGraph} \cite{Klashed:2010:UVU}, which facilitates \href{https://www.youtube.com/watch?v=e46cofM4lOs}{scale-independent rendering using scene graphs}. Axelsson et al. \cite{Axelsson:2017:DSG} later extended this concept to the \emph{Dynamic Scene Graph}, which, in the OpenSpace system \cite{Bladin:2018:GBC}, supports several high-detail locations and stereoscopic rendering. The Dynamic Scene Graph uses a dynamic camera node attachment to visualize scenes of varying scale and with high floating point precision.

\footnotetext{For example 25143 Itokawa, which was visited by the Hayabusa probe.}


With genome data we face similar problems concerning scale-dependent data and the need to traverse a range of scales. We also face the challenge that our conceptual scales are packed much more tightly in scale space as we explain next. This leads to fundamental differences between both application domains.

\section{Multi-Scale Genome Visualization}
\label{sec:concept}



Visualizing the nuclear human genome---from the nucleus that contains all chromosomal genetic material down to the very atoms that make up the DNA---is challenging due to the inherent organization of the DNA in tubular arrangements. DNA in its B-form is only \SI{2}{\nano\metre}\cite{Annunziato:2008:DPN} wide, which in its fibrous form or at more detailed scales would be too thin to be perceived. This situation is even more aggravated by the dense organization of the DNA and the structural hierarchy that bridges several scales. The previously discussed methods do not deal with such a combination of structural characteristics. Below we thus discuss the challenges that arise from the properties of these biological entities and how we address them by developing our new approach that smoothly transitions between views of the genome at its various scales.

\subsection{Challenges of interactive multiscale DNA visualization}
\label{sec:challenges}

Domain scientists who sequence, investigate, and generally work with genome data use a series of conceptual levels for analysis and visualization \cite{Nowotny:2015:GMO}: the \emph{genome} scale (containing all approx. 3.2\,Gb of the human genome), the \emph{chromosome} scale (50--100\,Mb), the \emph{loci} scale (in the order of Mb), the \emph{fiber} scale (in the order of Kb), the \emph{nucleosome} scale (146\,b), and the \emph{nucleotide} scale (\ie, 1\,b), in addition to the \emph{atomistic composition} of the nucleotides. These seven scales cover a range of approx.\ 4--5 orders of magnitude in physical size. In astronomy or astrophysics, in contrast, researchers deal with a similar number of scales:\footnote{Fu and Hanson \cite{Fu:2007:TSV} provide a nice overview in their Table~1.} approx.\ 7--8 conceptual scales of objects, yet over a range of some 24 orders of magnitude of physical size.\footnote{We only count explicit objects, not distances between objects. We also include smaller asteroids in the order of 10\textsuperscript{2}\,m. And we use the size of the observable universe at 2\,\texttimes\,$13.8\cdot10^{9}$\,light years $=2.6\cdot 10^{26}$\,m.} A fundamental difference between multi-scale visualizations in the two domains is, therefore, the \emphbf{scale density of the conceptual levels} that need to be depicted.

\begin{figure}[t]
	\centering
	\includegraphics[width=\columnwidth]{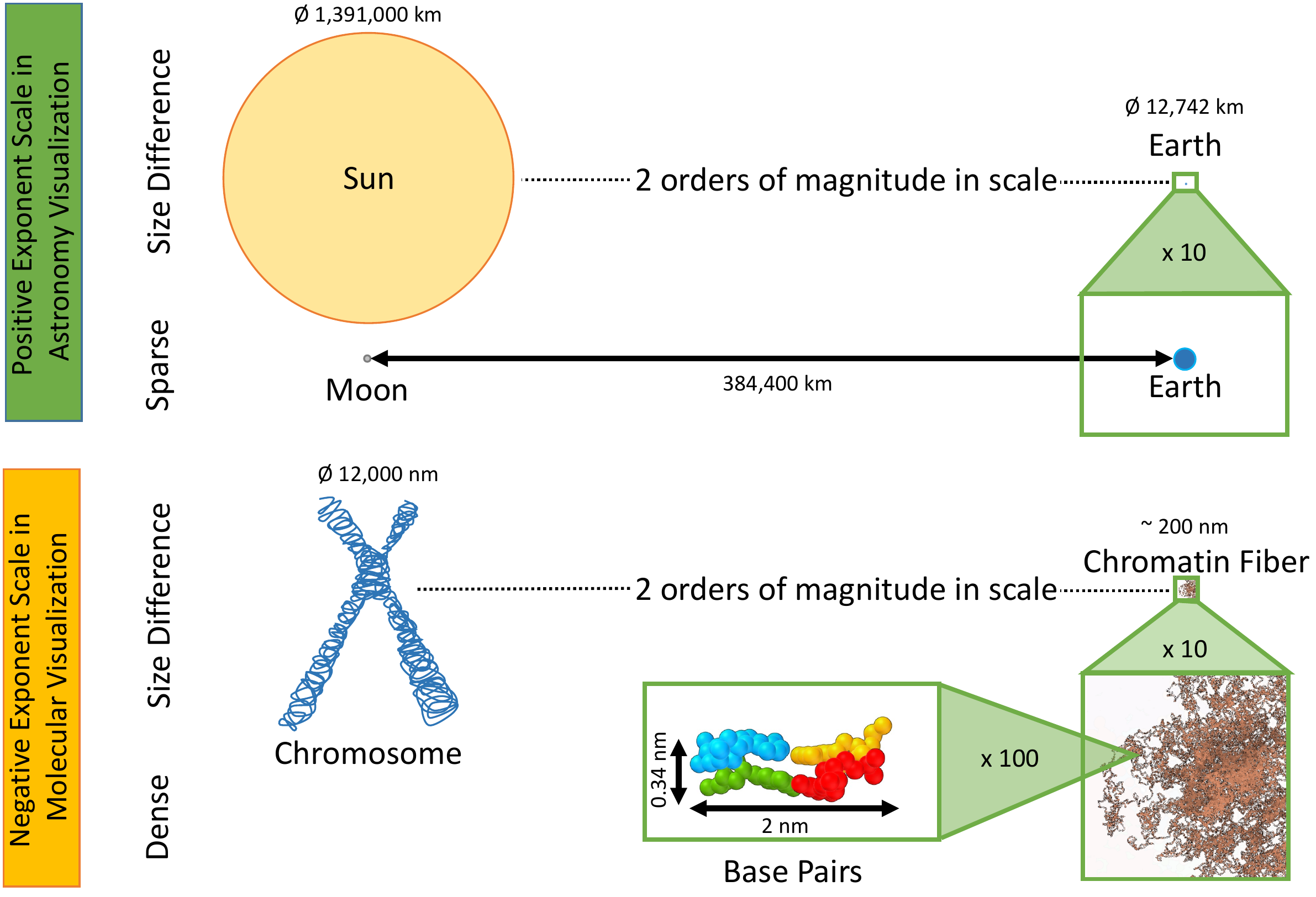}\vspace{-1ex}
	\caption{Multi-scale visualization in astronomy vs.\ genomics. The size difference between celestial bodies is extremely large (\eg, sun vs.\ earth---the earth is almost invisible at that scale). The distance between earth and moon is also large, compared to their sizes. In the genome, we have similar relative size differences, yet molecules are densely packed as exemplified by the two base pairs in the DNA double helix.}\vspace{-1ex}
	\label{fig:astronomyVsMolecular}
\end{figure}

\refstepcounter{footnote}
\footnotetext{Positive-exponent scale-space refers to measurement in meters, \ie, everything larger than approx.\ $1 \cdot 10^0$\,m.}
\refstepcounter{footnote}
\footnotetext{This means, everything of size approx. $1 \cdot 10^{-1}$\,m and smaller.}
\addtocounter{footnote}{-2}

Multi-scale astronomy visualization \cite{Klashed:2010:UVU,Axelsson:2017:DSG,Fu:2007:TSV,Hanson:2000:VLS} deals with \emphbf{positive-exponent scale-space}\footnotemark\ (\autoref{fig:astronomyVsMolecular}, top), where two neighboring scales are relatively far apart in scale space. For example, planets are much smaller than stars, stars are much smaller than galaxies, galaxies are much smaller than galaxy clusters, etc. On average, two scales have a distance of three or more orders of magnitude in physical space. The consequence of this high distance in scale space between neighboring conceptual levels is that, as one zooms out, elements from one scale typically all but disappear before the elements on the next conceptual level become visible. This aspect is used in creating multi-scale astronomy visualizations. For example, Axelsson et al.'s Dynamic Scene Graph \cite{Axelsson:2017:DSG} uses \emph{spheres of influence} to control the visibility range of objects from a given subtree of the scene graph.
In fact, the low scale density of the conceptual levels made the seamless animation of the astronomy/astrophysics section in \href{http://www.eamesoffice.com/the-work/powers-of-ten/}{the ``Powers of Ten'' Video} \cite{Eameses:1977:PTR} from 1977 possible---in a time before computer graphics could be used to create such animations. Eames and Eames \cite{Eameses:1977:PTR} simply and effectively blended smoothly between consecutive images that depicted the respective scales. For the cell/genome part, however, they use sudden transitions between conceptual scales without spatial continuity, and they also leave out several of the conceptual scales that scientists use today such as the chromosomes and the nucleosomes.


The reason for this problem of smoothly transitioning between scales in genome visualization---\ie, in \emphbf{\mbox{ne\-ga\-tive-ex\-po\-nent} \mbox{scale-space}}\footnotemark\ (\autoref{fig:astronomyVsMolecular}, bottom)---is that the conceptual levels of a multi-scale visualization are much closer to each other in scale. In contrast to astronomy's po\-si\-tive-ex\-po\-nent scale-space, there is only an average scale distance of about 0.5--0.6 orders of magnitude of physical space between two conceptual scales. Elements on one conceptual scale are thus still visible when elements from the next conceptual scale begin to appear. The scales for genome visualizations are thus much denser compared to astronomy's average scale distance of three orders of magnitude.

Moreover, in the genome the building blocks are \emphbf{physically connected in space and across conceptual scales}, except for the genome and chromosome levels. From the atoms to the chromosome scale, we have a single connected component. It is assembled in different geometric ways, depending on the conceptual scale at which we choose to observe. For example, the sequence of all nucleotides (base pairs) of the 46 chromosomes in a human cell would stretch for \SI{2}{\metre}, with each base pair only being \SI{2}{\nano\metre} wide \cite{Annunziato:2008:DPN}, while a complete set of chromosomes fits into the \SI{6}{\micro\metre} wide nucleus. Nonetheless, in all scales between the sequence of nucleotides and a chromosome we deal with the same, physically connected structure. In astronomy, instead, the physical space between elements within a conceptual scale is mostly empty and elements are physically not connected---elements are only connected by proximity (and gravity), not by visible links.

The large inter-scale distance and physical connectedness, naturally, also create the problem of how to visualize the \emphbf{relationship between two conceptual scale levels}. The mentioned multi-scale visualization systems from astronomy \cite{Klashed:2010:UVU,Axelsson:2017:DSG,Fu:2007:TSV,Hanson:2000:VLS} use animation for this purpose, sometimes adding invisible and intangible elements such as orbits of celestial bodies. In general multi-scale visualization approaches, \emph{multi-scale coordinate grids} \cite{Glueck:2009:MRV} can assist the perception of scale-level relationships. These approaches only work if the respective elements are independent of each other and can fade visually as one zooms out, for example, into the next-higher conceptual scale. The connected composition of the genome does make these approaches impossible.
In the genome, in addition, we have a \emphbf{complete model for the details in each conceptual level}, derived from data that are averages of measurements from many experiments on a single organism type. We are thus able to and need to show visual detail everywhere---as opposed to only close to a single point like planet Earth in astronomy.

Ultimately, all these points lead to \textbf{two fundamental challenges} for us to solve. The first (discussed in \autoref{sec:visual-embedding} and~\ref{sec:scale-dependent-camera}) is how to \emph{visually} create effective transitions between conceptual scales. The transitional scales shall show the containment and relationship character of the data even in still images and seamlessly allow us to travel across the scales as we are interacting. They must deal with the continuous nature of the depicted elements, which are physically connected in space and across scales. The second challenge is a computational one. Positional information of all atoms from the entire genome would not fit into GPU memory and will prohibit interactive rendering performance. We discuss how to overcome these computational issues in \autoref{sec:implement}, along with the implementation of the visual design from \autoref{sec:visual-embedding} and~\ref{sec:scale-dependent-camera}.

\subsection{Visual embedding of conceptual scales}
\label{sec:visual-embedding}

\new{Existing multi-scale visualizations of DNA \cite{Zwan:2011:IMV,Lueks:2011:SCC,Miao:2018:MVS} or other data \cite{Mohammed:2018:AVT} often use geometry manipulations to transition from one scale to the next. For the full genome, however, this approach would create too much detail to be useful and would require too many elements to be rendered. Moreover, two consecutive scales may differ significantly in structure and organization. A nucleosome, \eg, consists of nucleotides in double-helix form, wrapped around a histone protein. We thus need appropriate \emph{abstracted representations} for the whole set of geometry in a given scale that best depict the scale-dependent structure and still allow us to create smooth transitions between scales.}

\newlength{\pictureheight}
\setlength{\pictureheight}{0.3\columnwidth}
\begin{figure}[t]
  \centering
  \includegraphics[height=\pictureheight]{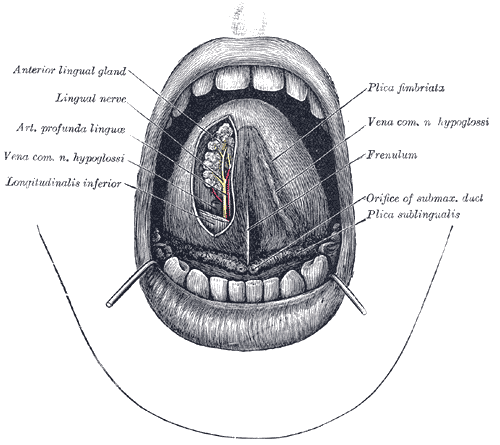}\hfill
  \includegraphics[height=\pictureheight]{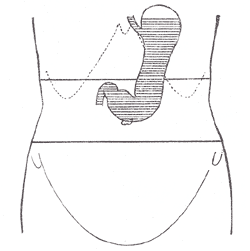}\hfill
  \includegraphics[height=\pictureheight]{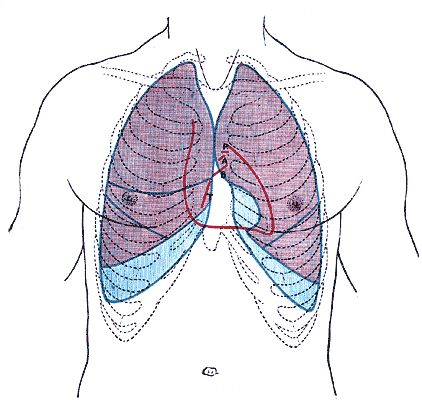}\vspace{-1ex}
  \caption{\href{https://commons.wikimedia.org/wiki/Gray's_Anatomy_plates}{Plates 1013, 1048, and 1216 from \emph{Gray's Anatomy}} \cite{Gray:1918:AHB}, demonstrate layered composition of multi-scale 3D objects by traditional illustrators. The images are in the public domain \ccPublicDomain.}\vspace{-1ex}
  \label{fig:grays_anatomy}
\end{figure}

Nonetheless, the mentioned geometry-based multi-scale transformations still serve as an important inspiration to our work. They often provide intermediate representations that may not be entirely accurate, but show how one scale relates to another one, even in a still image. Viewers can appreciate the properties of both involved scale levels, such as in Miao et al.'s \cite{Miao:2018:MVS} transition between nucleotides and strands.
Specifically, we take inspiration from traditional illustration where a related visual metaphor has been used before. As exemplified by \autoref{fig:grays_anatomy}, illustrators sometimes use an abstracted representation of a coarser scale to aid viewers with understanding the overall composition as well as the spatial location of the finer details. This embedding of one representation scale into the next is similar to combining several layers of visual information---or super-imposition \cite[pp.~288\,ff]{Munzner:2014:VAD}. It is a common approach, for example, in creating maps. In visualization, this principle has been used in the past (\eg, \cite{Interrante:1996:ITC,Rautek:2007:SLI,Carnecky:2012:MID,Rocha:2017:DRL}), typically applying some form of transparency to be able to perceive the different layers.
Transparency, however, can easily lead to visualizations that are difficult to understand \cite{Carnecky:2013:STI}. Simple outlines to indicate the coarser shape or context can also be useful \cite{Tietjen:2005:CSS}. In our case, even outlines easily lead to clutter due to the immense amount of detail in the genome data. Moreover, we are not interested in showing that some elements are \emph{spatially} inside others, but rather that the elements are \emph{part of a higher-level structure}, thus are \emph{conceptually} contained.


\begin{figure}[t!]
	\centering
	\includegraphics[width=\columnwidth]{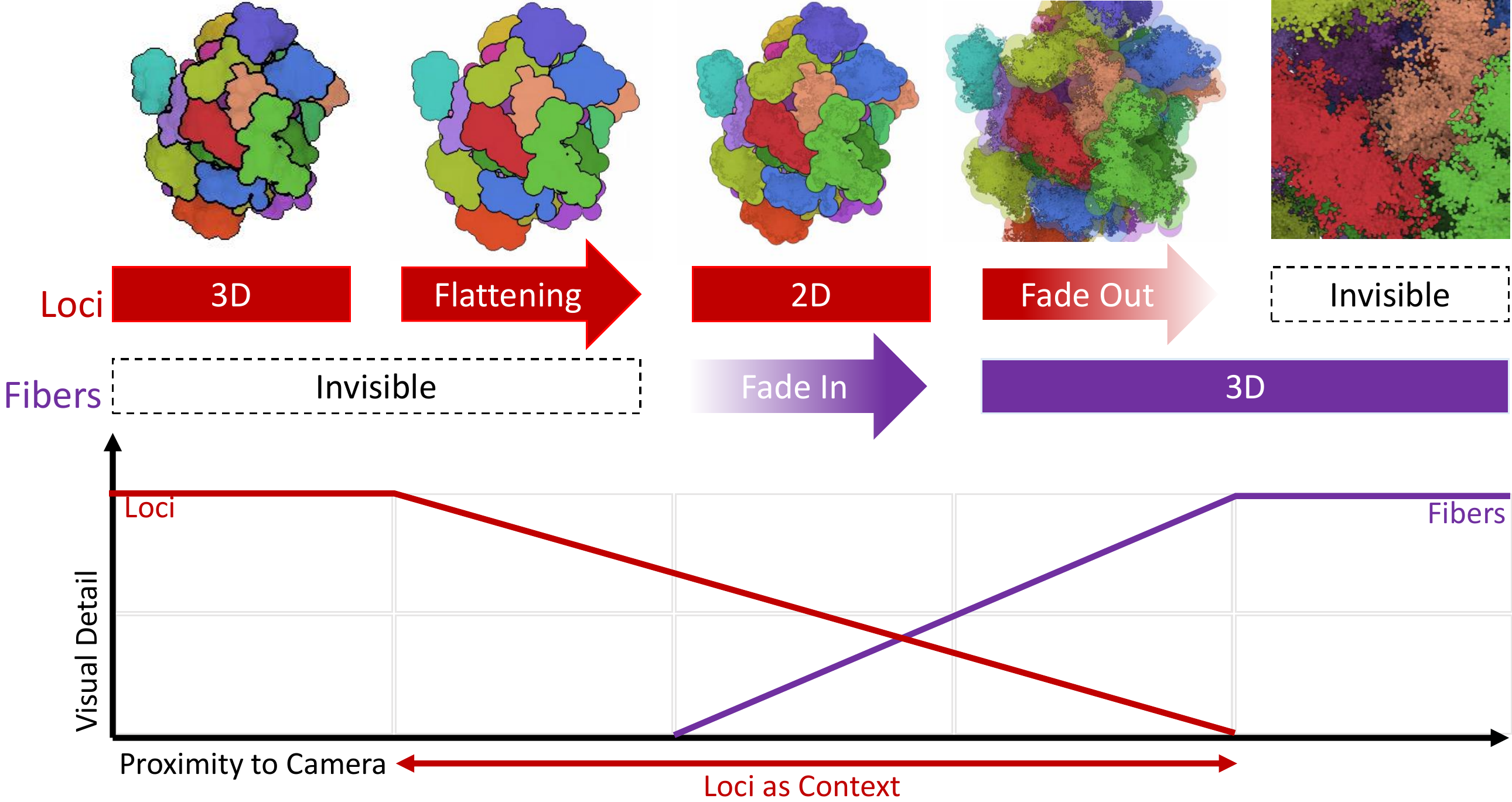}\vspace{-3ex}
	\caption{Visual embedding, schematic principle.}\vspace{-1ex}
	\label{fig:visual_embedding}
\end{figure}

\begin{figure}[t]
	\centering
	\setlength{\subfigcapskip}{-2.5ex}%
	\subfigure[\hspace{.535\columnwidth}]{\label{fig:canvas:a}\includegraphics[height=.43\columnwidth]{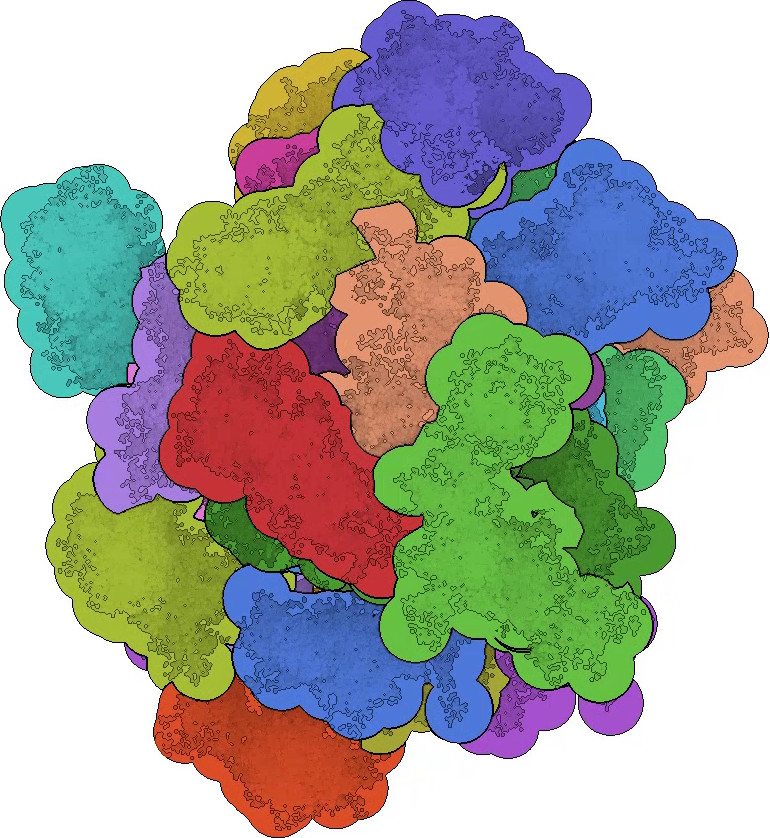}}\hfill%
	\subfigure[\hspace{.535\columnwidth}]{\label{fig:canvas:b}\includegraphics[height=.43\columnwidth]{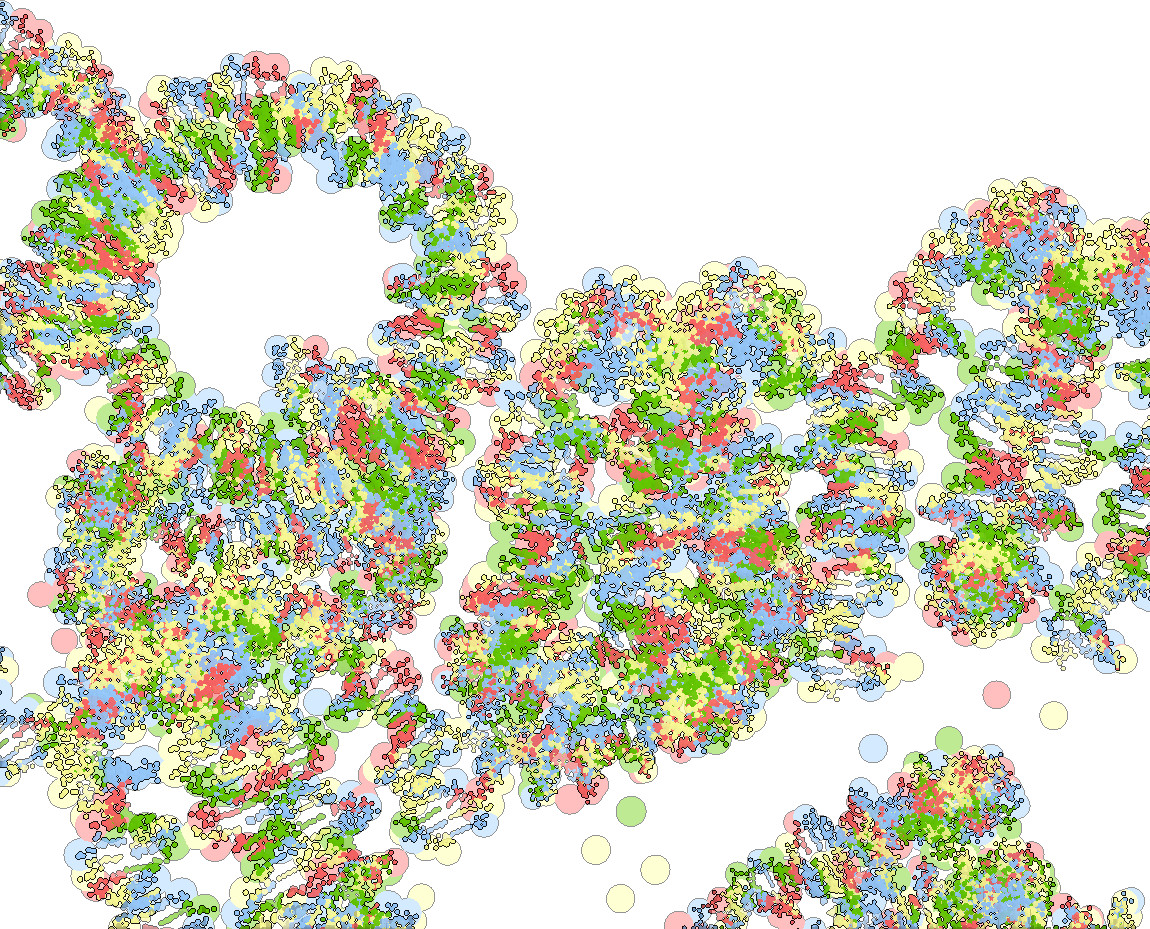}}\vspace{-1ex}
	\caption{Two snapshots of scale transition views, \subref{fig:canvas:a} between the chromosome and the detailed chromosome scales, as well as \subref{fig:canvas:b} between nucleotides and detailed nucleotides scales.}\vspace{-1ex}
	\label{fig:canvas}
\end{figure}

We therefore propose \emphbf{visual scale embedding} of the detailed scale into its coarser parent (see the illustration in \autoref{fig:visual_embedding}). 
We render an abstracted representation of the coarser scale first to serve as a context or canvas, and render the representation of the more detailed scale on top of it. The context or canvas should not interfere in its spatial perception with the depiction of the detail because it is typically \emph{surrounding} the next scale. An exemplifying case of how this can lead to perception issues in still images was given by Svetachov et al.\ \cite[Fig. 10]{Svetachov:2010:DCI}. We thus completely flattened the context as shown in \autoref{fig:visual_embedding} and inspired by previous multi-scale visualizations from structural biology \cite{Parulek:2014:CLD}. Then we render the detailed geometry of the next-smaller scale on top of it. This concept adequately supports our goal of smooth scale transitions. A geometric representation of the coarser scale is first shown using 3D shading as long as it is still small on the screen, \ie, the camera is far away. It transitions to a flat, canvas-like representation when the camera comes closer and the detail in this scale is not enough anymore. We now add the representation of the more detailed scale on top---again using 3D shading, as shown for two scale transitions in \autoref{fig:canvas}.

Our illustrative visualization concept combines the 2D aspect of the flattened coarser scale with the 3D detail of the finer scale. With it we make use of superimposed representations as argued by Viola and Isenberg \cite{Viola:2018:PCA}, which are an alternative to spatially or temporally juxtaposed views. In our case, the increasingly abstract character of rendering of the coarser scale (as we flatten it during zooming in) relates to its increasingly contextual and conceptual nature. Our approach thus relates to \emph{semantic zooming} \cite{Perlin:1993:PAA} because the context layer turns into a flat surface or canvas, irrespective of the underlying 3D structure and regardless of the specific chosen view direction. This type of scale zoom does not have the character of cut-away techniques as often used in tools to explore containment in 3D data (\eg, \cite{Li:2007:ICI,LeMuzic:2016:VEC}). Instead, it is more akin to the semantic zooming in the visualization of abstract data, which is embedded in the 2D plane (\eg, \cite{Weaver:2004:BHC}).

\subsection[Multi-scale visual embedding and scale-dependent view]{\textls[-14]{Multi-scale visual embedding and scale-dependent view}}
\label{sec:scale-dependent-camera}

\begin{figure}[t!]
	\centering
	\includegraphics[width=.9\columnwidth]{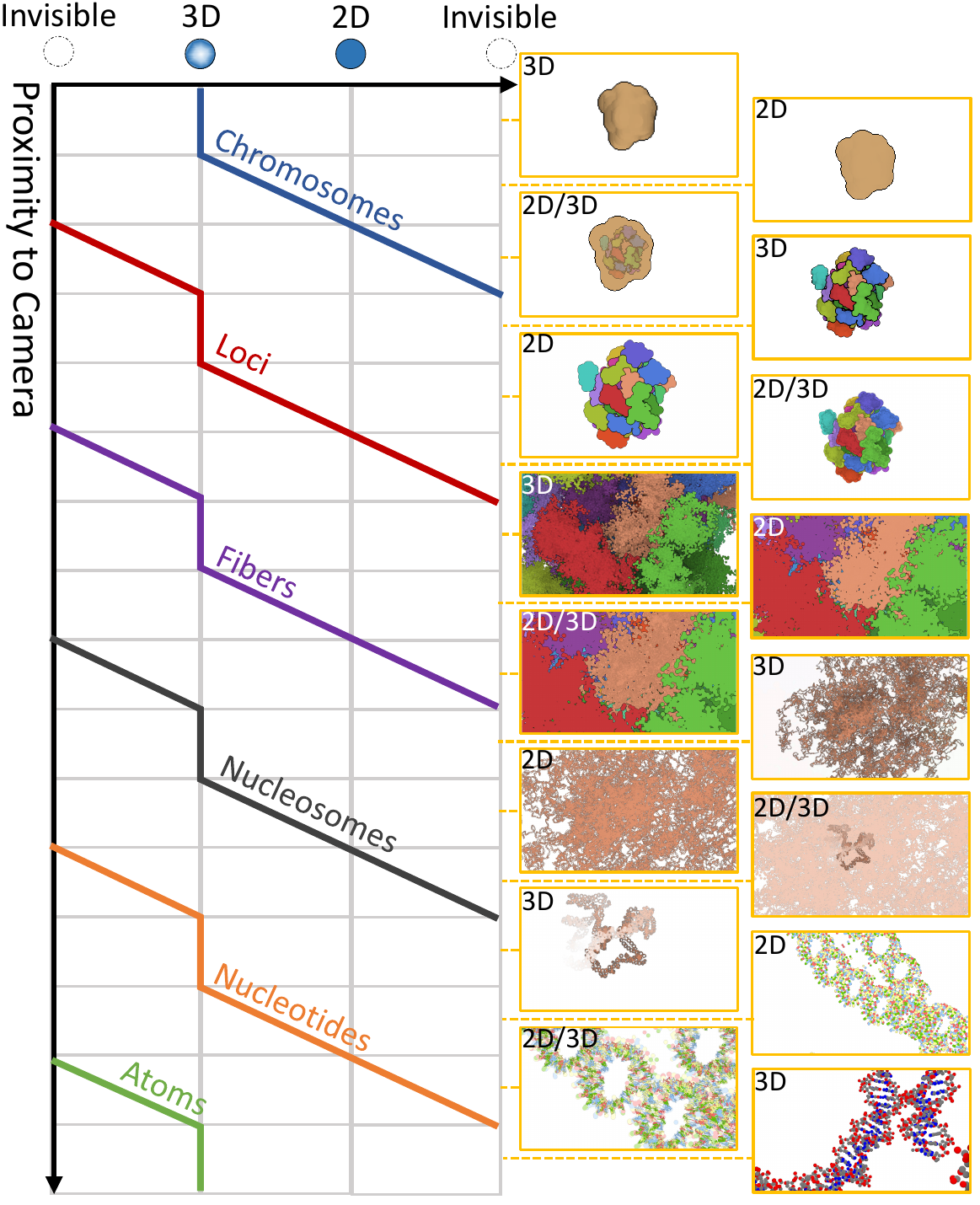}\vspace{-1ex}
	\caption{Sequence of visual scale embeddings, based on the data levels.}\vspace{-2ex}
	\label{fig:visual-embedding-hierarchy}
\end{figure}

One visual embedding step connects two consecutive semantic scales. We now concatenate several steps to assemble the whole hierarchy (\autoref{fig:visual-embedding-hierarchy}). This is conceptually straightforward because each scale by itself is shown using 3D shading. Nonetheless, as we get to finer and finer details, we face the two major problems mentioned at the start of \autoref{sec:visual-embedding}: visual clutter and limitations of graphics processing. Both are caused by the tight scale space packing of the semantic levels in the genome. At detailed scales, a huge number of elements are potentially visible, \eg, 3.2\,Gb at the level of nucleotides. To address this issue, we adjust the camera concept to the multi-scale nature of the data.

In previous multi-scale visualization frameworks \cite{Hanson:2000:VLS,Fu:2007:TSV,Klashed:2010:UVU,Axelsson:2017:DSG}, researchers have already used \emphbf{scale-constrained camera navigation}. For example, they apply a \emphbf{scale-dependent camera speed} to quickly cover the huge distances at coarse levels and provide fine control for detailed levels. In addition, they used a \emphbf{scale-dependent physical camera size or scope} such that the depicted elements would appropriately fill the distance between near and far plane, or use depth buffer remapping \cite{Fu:2007:TSV} to cover a larger depth range. In astronomy and astrophysics, however, we do not face the problem of a lot of nearby elements in detailed levels of scale due to their loose scale-space packing. After all, if we look into the night sky we do not see much more than ``a few'' stars from our galactic neighborhood which, in a visualization system, can easily be represented by a texture map. Axelsson et al. \cite{Axelsson:2017:DSG}, for example, simply attach their cameras to nodes within the scale level they want to depict.

For the visualization of genome data, however, we have to introduce an active control of the \emphbf{scale-dependent \emph{data-hierarchy} size or scope} as we would ``physically see,'' for example, all nucleosomes or nucleotides up to the end of the nucleus. Aside from the resulting clutter, such \new{complete genome} views would also \emph{conceptually} not be helpful because, due to the nature of the genome, the elements within a detailed scale largely repeat themselves. The visual goal should thus be to only show a relevant and scale-dependent subset of each hierarchy level. We thus limit the rendering scope to a subset of the hierarchy, depending on the chosen scale level and spatial focus point. The example in \autoref{fig:scale-dependent-hierarchy-scope} depicts the nucleosome scale, where we only show a limited number of nucleosomes to the left and the right of the current focus point in the sequence, while the rest of the hierarchy has been blended out. We thereby extend the visual metaphor of the canvas, which we applied in the visual embedding, and use the white background of the frame buffer as a second, \emphbf{scale-dependent canvas}, which limits the visibility of the detail. In contrast to photorealism\footnote{Of course, visualizations in astronomy also comprise non-photographic components such as hyperspectral imaging or radio astronomy data, but many scales use depictions based on a photographic camera as their guiding vision.} that drives many multi-scale visualizations in astronomy, we are interested in appropriately abstracted representations through a \emphbf{scale-dependent removal of distant detail} to support viewers in focusing on their current region of interest.

\section{Implementation}
\label{sec:implement}

Based on the conceptual design from \autoref{sec:concept} we now describe the implementation of our multi-scale genome visualization framework. We first describe the used and then explain the shader-based realization of the scale transitions using a series of visual embedding steps as well as some interaction considerations.

\subsection{Data sources and data hierarchy}
\label{sec:implementation:data}

Researchers in genome studies have a high interest in understanding the relationships between the spatial structure at the various scale levels and the biological function of the DNA. Therefore they have created a multi-scale dataset that allows them to look at the genome in different spatial scale levels \cite{Nowotny:2015:GMO}. This data was derived by Nowotny et al.\ \cite{Nowotny:2015:GMO} from a model of the human genome by Asbury et al.\ \cite{Asbury:2010:G3D}, which in turn was constructed based on various data sources and observed properties. For determining the positions of the chromatin fiber, Nowotny et al.\ used Bancaud et al.'s \cite{Bancaud:2012:FMN} approach of space-filling, fractal packing. As a result, Nowotny et al.\ \cite{Nowotny:2015:GMO} obtained the positions of the \emph{nucleotides} in space, and from these computed the positions of \emph{fibers}, \emph{loci}, and \emph{chromosomes} (\autoref{fig:nowotny}). They stored this data in their own Genome Scale System (GSS) format and also provided the positions of the \emph{nucleotides} for one nucleosome (\autoref{fig:nowotny}, bottom-right). Even with this additional data, we still have to procedurally generate further information as we visualize this data such as the orientations of the nucleosomes \new{(based on the location of two consecutive nucleosomes)} and the \new{linker DNA} strands of nucleotides connecting two consecutive nucleosomes.

\begin{figure}[t!]
	\centering
	\vspace{-1ex}\includegraphics[width=\columnwidth]{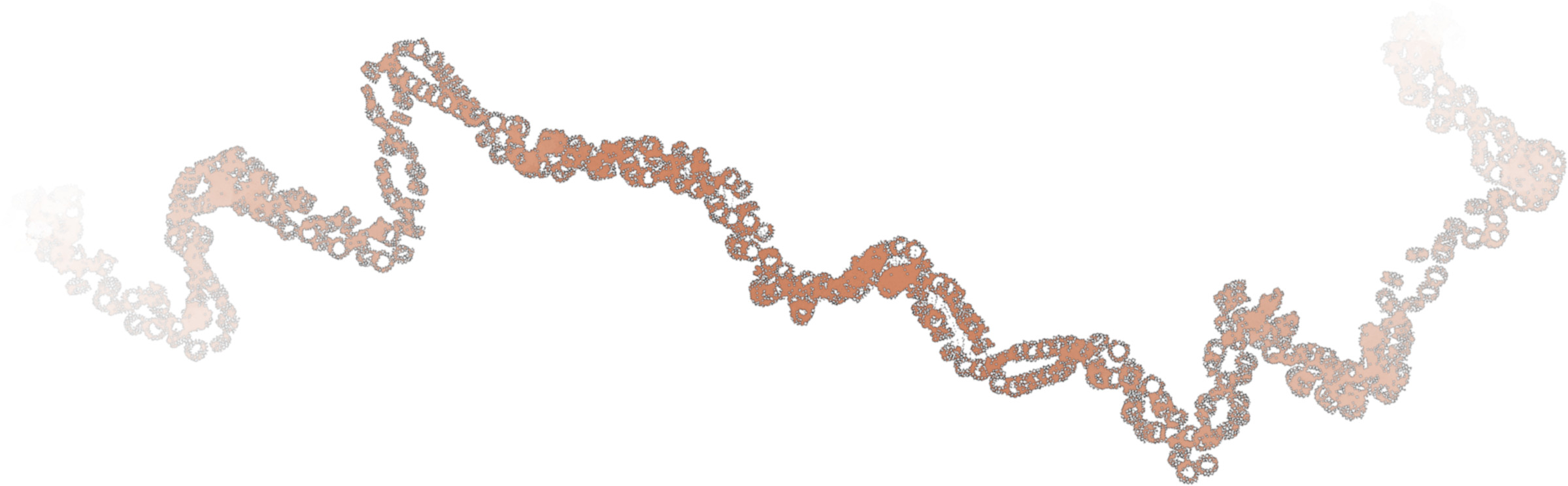}\vspace{-2ex}
	\caption{Scale-dependent hierarchy scope realized for nucleosomes by showing five \emph{fiber} locations around the focus and fading out the ends.}\vspace{-.5ex}
	\label{fig:scale-dependent-hierarchy-scope}
\end{figure}

\begin{figure}[t!]
	\centering
	\includegraphics[height=.23\columnwidth,trim={0 1631px 1112px 0},clip]{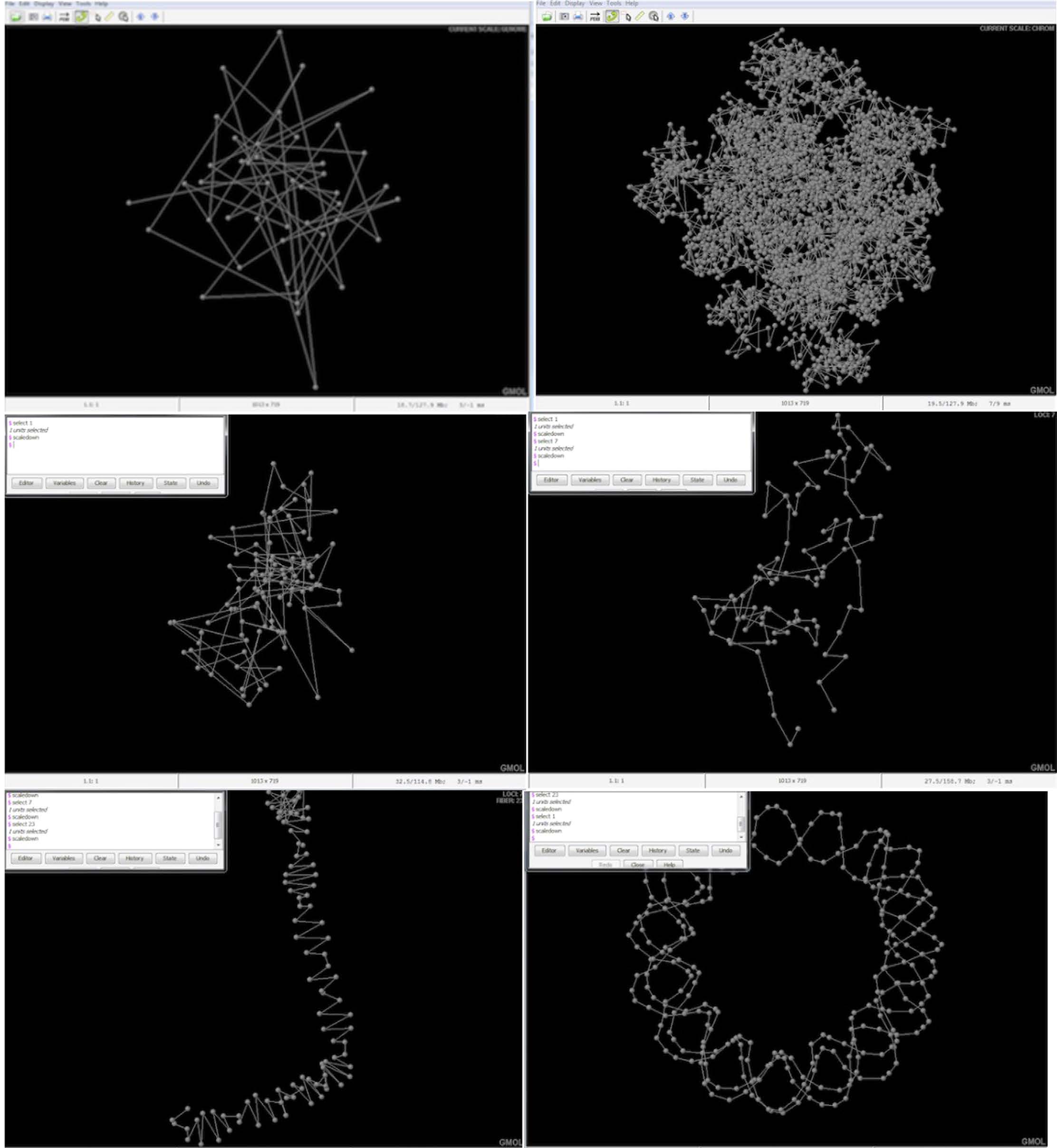}\hfill%
	\includegraphics[height=.23\columnwidth,trim={1112px 1631px 0 0},clip]{nowotny}\hfill%
	\includegraphics[height=.23\columnwidth,trim={0 790px 1112px 841px},clip]{nowotny}\hfill\\
	\includegraphics[height=.23\columnwidth,trim={1112px 795px 0 836px},clip]{nowotny}\hfill%
	\includegraphics[height=.23\columnwidth,trim={0 0 1112px 1631px},clip]{nowotny}\hfill%
	\includegraphics[height=.23\columnwidth,trim={1112px 0 0 1631px},clip]{nowotny}\vspace{-1ex}%
	\caption{Screenshots from GMOL showing traditional visualizations of the multi-scale genome data by depicting chromosomes, loci of all chromosomes, loci of a single chromosome, fibers, nucleosomes, and nucleotides. Images from Nowotny et al. \cite{Nowotny:2015:GMO} \href{https://creativecommons.org/licenses/by/4.0/}{(\ccLogo\ \ccAttribution\ CC BY 4.0)}.}\vspace{-1.5ex}
	\label{fig:nowotny}
\end{figure}

\begin{figure}[t!]
	\centering
	\setlength{\subfigcapskip}{-2.5ex}%
	\subfigure[\hspace{.535\columnwidth}]{\label{fig:hierarchy-confusion:a}\includegraphics[height=0.32\columnwidth]{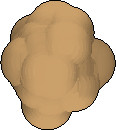}}\hfill%
	\subfigure[\hspace{.535\columnwidth}]{\label{fig:hierarchy-confusion:b}\includegraphics[height=0.32\columnwidth]{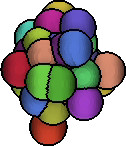}}\hfill%
	\subfigure[\hspace{.535\columnwidth}]{\label{fig:hierarchy-confusion:c}\includegraphics[height=0.32\columnwidth]{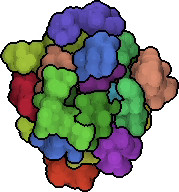}}%
	\caption{Mismatch of the data scale and the semantic scale: the \emph{chromosome} locations are not easily recognized as chromosomes in \subref{fig:hierarchy-confusion:b}, instead we display them using a single color as in \subref{fig:hierarchy-confusion:a} to represent the semantic \emph{nucleus} scale. The chromosomes and their three-dimensional shape are better shown using data from the more detailed \emph{loci} scale \subref{fig:hierarchy-confusion:c}.%
}\vspace{-2ex}
	\label{fig:hierarchy-confusion}
\end{figure}

This data provides positions at every scale level, without additional information about the actual sizes. Only at the nucleotide and atom scales the sizes are known. It was commonly thought that nucleosomes are tightly and homogeneously packed into \SI{30}{\nano\metre} fibers, \SI{120}{\nano\meter} chromonema, and 300--\SI{700}{\nano\meter} chromatids, but recent studies \cite{Ou:2017:CV3} disprove this organization and confirm the existence of flexible chains with diameters of 5--\SI{24}{\nano\meter}. Therefore, for all hierarchically organized scales coarser than the nucleosome, \new{we do not have information about the specific shape that each data point represents. We use spheres with scale-adjusted sizes as rendering primitives as they well portray the chaining of elements according to the data-point sequence.}


\begin{table}[t]
	\centering
	\scriptsize
	\caption{Relationship between data and semantic hierarchies.}
	\label{tab:hierarchy-mismatch}
	\begin{tabu}{@{}X[1.4,c,m]X[c,m]X[c,m]X[c,m]X[c,m]X[c,m]@{}}
		\toprule
		image & data level & colored by & elements rendered & semantic level & transition \\
		\midrule
		\setlength{\fboxsep}{2pt}
		\setlength{\fboxrule}{0pt}
		\fbox{\includegraphics[width=12mm]{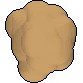}} & \emph{chromo\-some} positions & one single color & all chromo\-somes & nucleus & \raisebox{-2.5em}[0pt][0pt]{\rotatebox[origin=c]{270}{\large$\curvearrowright$}\begin{minipage}{10mm}\centering visual embedding\end{minipage}} \\
		\setlength{\fboxsep}{2pt}
		\setlength{\fboxrule}{0pt}
		\fbox{\includegraphics[width=12mm]{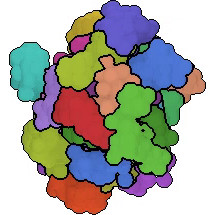}} & \emph{loci} positions & chromo\-some & all chromo\-somes & chromo\-some & \raisebox{-2.5em}[0pt][0pt]{\rotatebox[origin=c]{270}{\large$\curvearrowright$}\begin{minipage}{10mm}\centering visual embedding\end{minipage}} \\
		\setlength{\fboxsep}{2pt}
		\setlength{\fboxrule}{0pt}
		\fbox{\includegraphics[width=12mm]{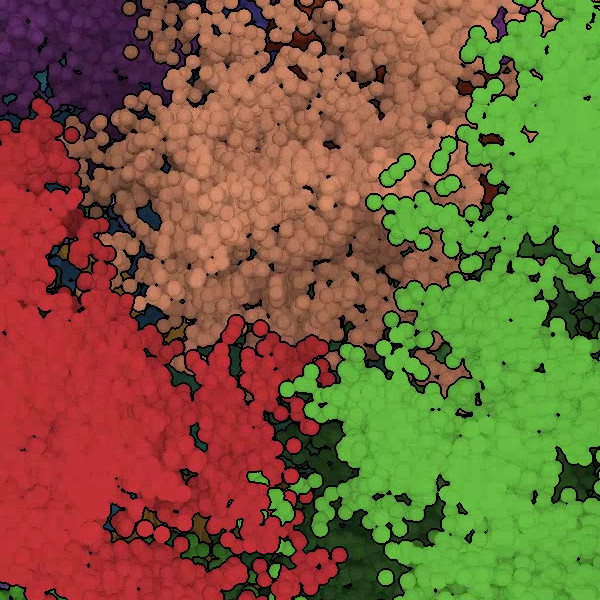}} & \emph{fiber} positions & chromo\-some & all chromo\-somes & chromo\-some with detail & \raisebox{-2.5em}[0pt][0pt]{\rotatebox[origin=c]{270}{\large$\curvearrowright$}\begin{minipage}{10mm}\centering visual embedding\end{minipage}} \\
		\setlength{\fboxsep}{2pt}
		\setlength{\fboxrule}{0pt}
		\fbox{\includegraphics[width=12mm]{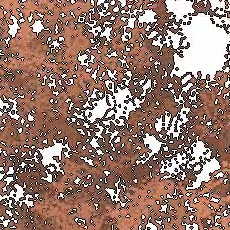}} & \emph{nucleo\-some} positions & chromo\-some & focus chromo\-some & fibers & \raisebox{-2.5em}[0pt][0pt]{\rotatebox[origin=c]{270}{\large$\curvearrowright$}\begin{minipage}{10mm}\centering visual embedding\end{minipage}} \\
		\setlength{\fboxsep}{2pt}
		\setlength{\fboxrule}{0pt}
		\fbox{\includegraphics[width=12mm]{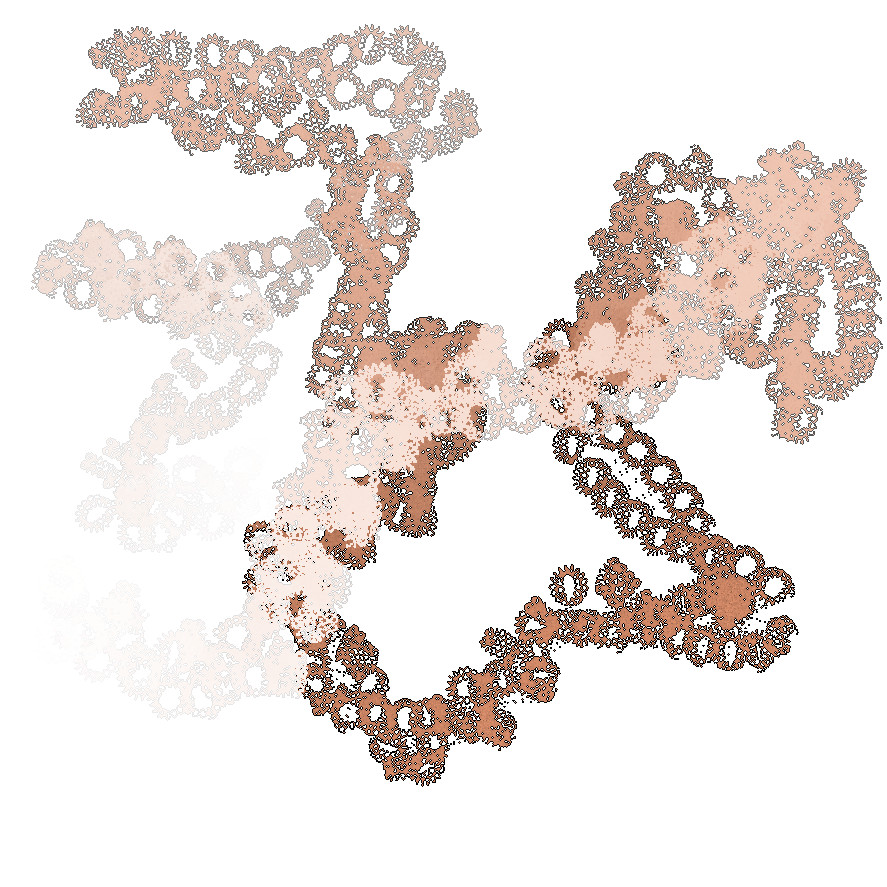}} & \emph{nucleo\-tide} positions & chromo\-some & nucleo\-somes of 5 fibers, with links & nucleo\-somes & \raisebox{-2.5em}[0pt][0pt]{\rotatebox[origin=c]{270}{\large$\curvearrowright$}\begin{minipage}{10mm}\centering color change\end{minipage}} \\
		\setlength{\fboxsep}{2pt}
		\setlength{\fboxrule}{0pt}
		\fbox{\includegraphics[width=12mm]{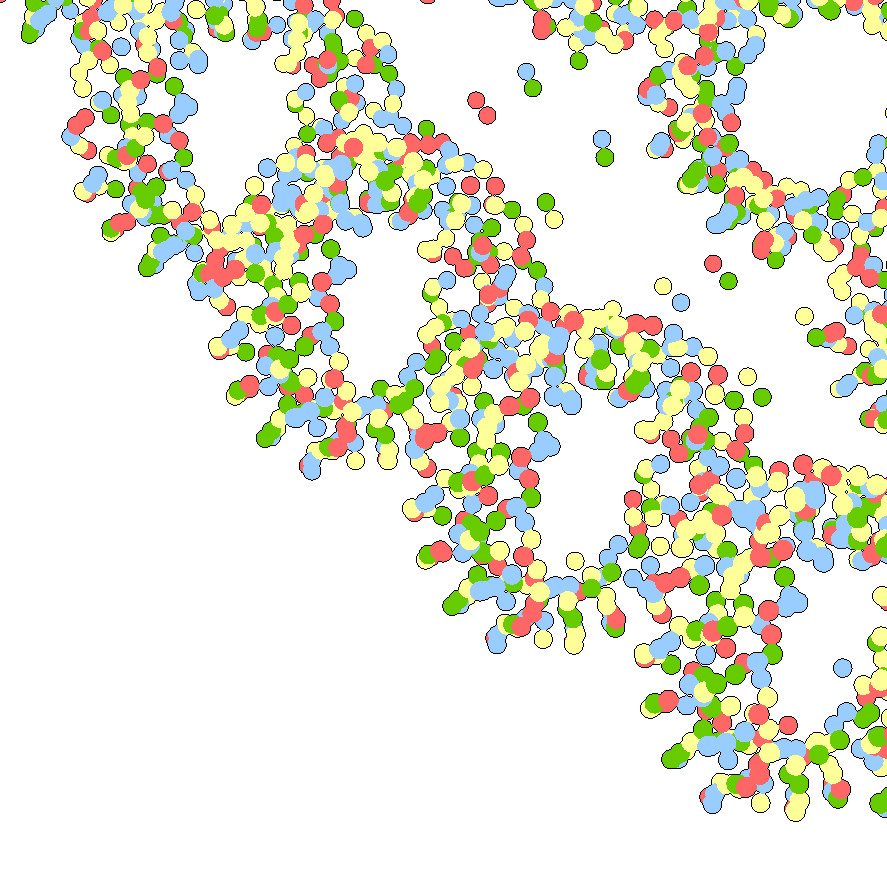}} & \emph{nucleo\-tide} positions & nucleo\-tide & nucleo\-somes of 5 fibers & nucleo\-tides & \raisebox{-2.5em}[0pt][0pt]{\rotatebox[origin=c]{270}{\large$\curvearrowright$}\begin{minipage}{10mm}\centering visual embedding\end{minipage}} \\
		\setlength{\fboxsep}{2pt}
		\setlength{\fboxrule}{0pt}
		\fbox{\includegraphics[width=12mm]{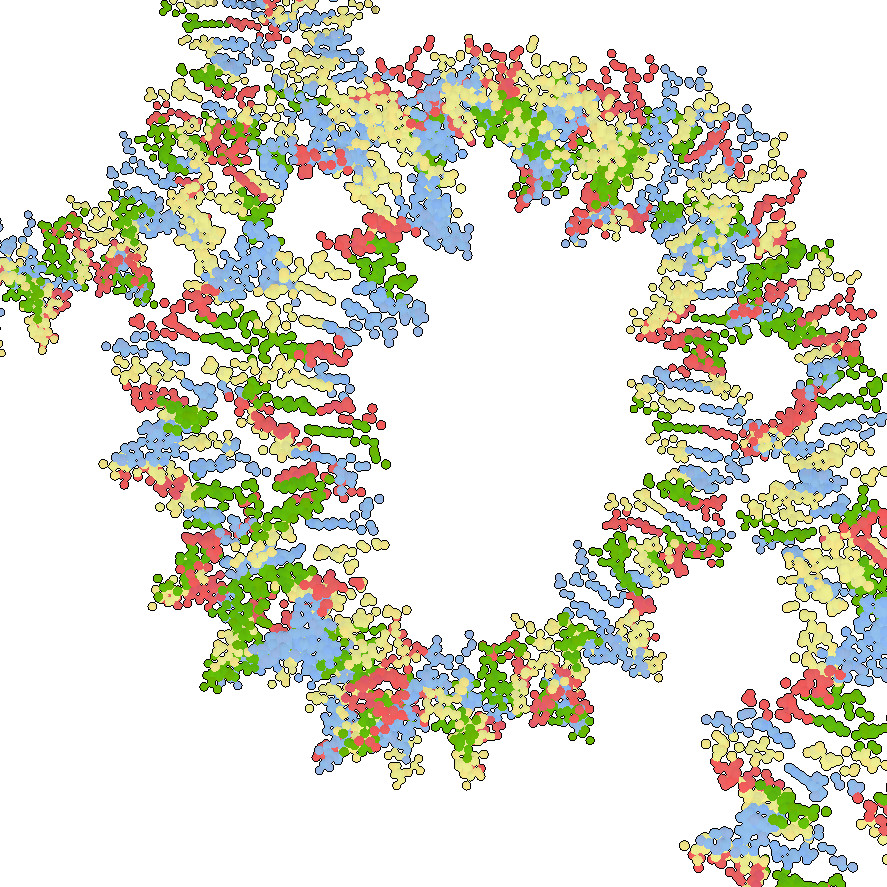}} & \emph{atom} positions & nucleo\-tide & nucleo\-somes of 5 fibers & nucleo\-tides with detail & \raisebox{-2.5em}[0pt][0pt]{\rotatebox[origin=c]{270}{\large$\curvearrowright$}\begin{minipage}{10mm}\centering color change\end{minipage}} \\
		\setlength{\fboxsep}{2pt}
		\setlength{\fboxrule}{0pt}
		\fbox{\includegraphics[width=12mm]{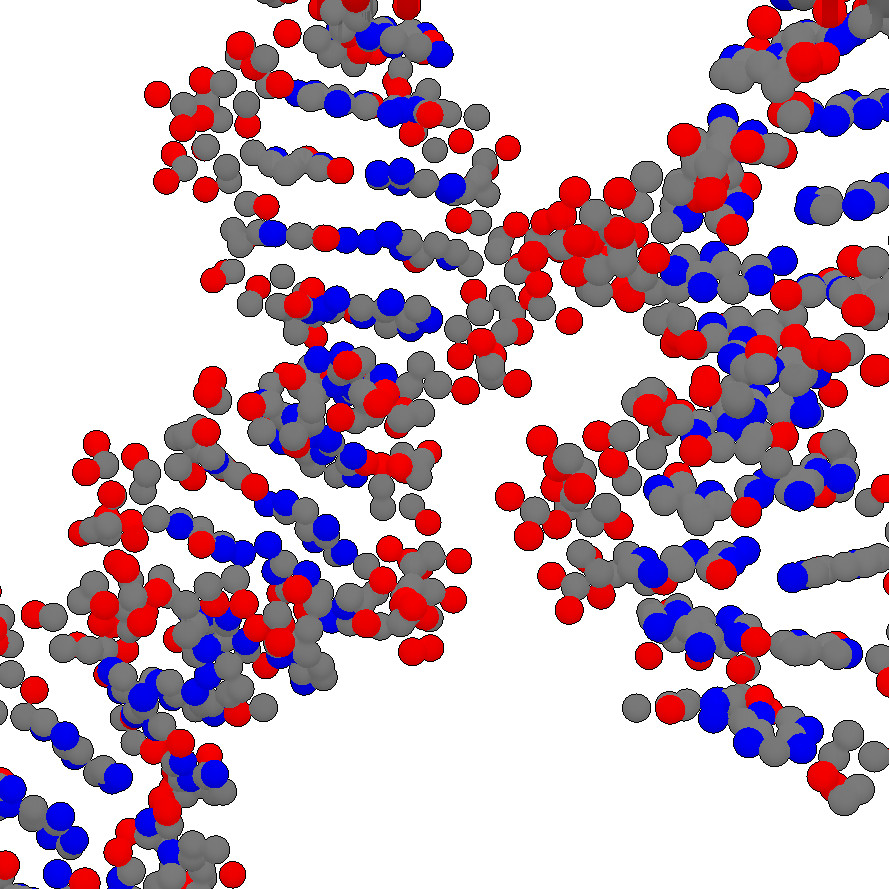}} & \emph{atom} positions & element & nucleo\-somes of 5 fibers & individual atoms \\
		\bottomrule
	\end{tabu}\vspace{-2ex}
\end{table}

With respect to visualizing this multi-scale phenomenon, the data hierarchy (\ie, 100 nucleosomes $=$ 1 fiber, 100 fibers $=$ 1 locus, approx.\ 100 loci $=$ 1 chromosome) is not the same as the hierarchy of semantic scales that a viewer sees. For example, the dataset contains a level that stores the \emph{chromosome} positions, but if rendered we would only see one sphere for each chromosome (\autoref{fig:hierarchy-confusion:b}). Such a depiction would not easily be recognized as representing a chromosome due to the lack of detail. The chromosomes by themselves only become apparent once we display them with more shape details using the data level of the \emph{loci} as given in \autoref{fig:hierarchy-confusion:c}. The locations at the chromosomes data scale can instead be better used to represent the semantic level of the \emph{nucleus} by rendering them as larger spheres, all with the same color and with a single outline around the entire shape as illustrated in \autoref{fig:hierarchy-confusion:a}.

In \autoref{tab:hierarchy-mismatch} we list the relationships between data hierarchy and semantic hierarchy for the entire set of scales we support. From the table follows that the choice of color assignment and the subset of rendered elements on the screen supports viewers in understanding the semantic level, which we want to portray. For example, by rendering the \emph{fiber} positions colored by \emph{chromosome} we facilitate the understanding of a detailed depiction of a chromosome, rather than that chromosomes consist of several loci. In an alternative depiction for domain experts, who are interested in studying the loci regions, we could instead assign the colors by \emph{loci} for the \emph{fiber} data level and beyond.

We added two additional scale transitions that are not realized by visual embedding, but instead by color transitions. The first of these transitions changes the colors from the previously maintained chromosome color to nucleotide colors as the \emph{nucleotide} positions are rendered in their 3D shape to illustrate that the \emph{nucleosomes} themselves consist of pairs of \emph{nucleotides}. The following transition then uses visual embedding as before, to transition to \emph{atoms} while maintaining nucleotide colors. The last transition, again changes this color assignment such that the atoms are rendered in their typical element colors, using 3D shading and without flattening them.

\subsection{Realizing visual scale embedding}

For our proof-of-concept implementation we build on the molecular visualization functionality provided in the Marion framework \cite{Mindek:2018:VMP}. We added to this framework the capability to load the previously described GSS data. We thus load and store the highest detail of the data---the 23,958,240 nucleosome positions---as well as all positions of the coarser scales. To show more detail, we use the single nucleosome example in the data, which consists of 292 nucleotides and then create the \mbox{$\approx 24 \cdot 10^6$}~instances for the semantic nucleosome scale. Here we fully use of Le Muzic et al.'s \cite{LeMuzic:2015:CEL} technique of employing the tessellation stages on the GPU, which dynamically injects the atoms of the nucleosome. We apply a similar instancing approach for transitioning to an atomistic representation, based on the \href{https://www.rcsb.org/structure/1AOI}{1AOI model from the PDB}. To visually represent the elements, we utilize 2D sphere impostors instead of sphere meshes \cite{LeMuzic:2015:CEL}. Specifically, we use triangular 2D billboards (\ie, only three vertices) that always face the camera and assign the depth to each fragment that it would get if it had been a sphere.

If we wanted to directly render all atoms at the finest detail scale, we would have to deal with $\approx$\,3.2\,Gb\,$\cdot 70$\,atoms/b\,$=224 \cdot 10^9$ atoms. This amount of detail is not possible to render at interactive rates. With \ac{LOD} optimizations, such as the creation of super-atoms for distant elements, cell\-VIEW could process $15 \cdot 10^9$~atoms at 60\,Hz \cite{LeMuzic:2015:CEL}. This amount of detail does not seem to be necessary in our case. Our main goal is the depiction of the scale transitions and too much detail would cause visual noise and distractions. We use the \emph{scale-dependent removal of distant detail} described in \autoref{sec:scale-dependent-camera}. As listed in \autoref{tab:hierarchy-mismatch}, for coarse scales we show all chromosomes. Starting with the semantic fibers scale, we only show the focus chromosome. For the semantic nucleosomes level, we only show the focus fiber and two additional fibers in both directions of the sequence. To indicate that the sequence continues, we gradually fade out the ends of the sequence of nucleosomes as shown in \autoref{fig:scale-dependent-hierarchy-scope}. For finer scales beyond the nucleosomes, we maintain the sequence of five fibers around the focus point, but remove the detail of the links between nucleosomes.

To manage the different rendering scopes and color assignments, we assign IDs to elements in a data scale and record the IDs of the hierarchy ancestors of an element. For example, each \emph{chromosome} data element gets an ID, which in turn is known to the \emph{loci} data instances. We use this ID to assign a color to the chromosomes. Because we continue rendering all chromosomes even at the \emph{fiber} data level respectively semantic \emph{chromosome with detail} level, we also pass the IDs of the chromosomes to the \emph{fiber} data elements. Later, the IDs of the \emph{fiber} data elements are used to determine the rendering scope in the data levels of \emph{nucleotide} positions and finer (more detail).

For realizing the transition in the \emph{visual scale embedding}, \ie, transitioning from the coarser scale $S_N$ to the finer scale $S_{N+1}$, we begin by alpha-blending $S_N$ rendered with 3D detail and flattened $S_N$. We achieve the 3D detail with screen-space ambient occlusion (SSAO), while the flattened version does not use SSAO. Next we transition between $S_N$ and $S_{N+1}$ by first rendering $S_N$ and then $S_{N+1}$ on top, the latter with increasing opacity. Here we avoid visual clutter by only adding detail to elements in $S_{N+1}$ on top of those regions that belonged to their parents in $S_N$. The necessary information for this purpose comes from the previously mentioned IDs. We thus first render all flattened elements of $S_N$, before blending in detail elements from $S_{N+1}$. In the final transition of \emph{visual scale embedding}, we remove the elements from $S_N$ through alpha-blending. For the two color transitions discussed in \autoref{sec:implementation:data} we simply alpha-blend between the corresponding elements of $S_N$ and $S_{N+1}$, but with different color assignments.

\subsection{Interaction considerations}
\label{sec:interaction}

\new{The rendering speeds are in the range of 15--35\,fps on an Intel Core\textsuperscript{\texttrademark} PC (i7-8700K, 6 cores, 32 GB RAM, 3.70\,GHz, nVidia Quadro P4000, Windows 10 x64).} In addition to providing a scale-controlled traversal of the scale hierarchy toward a focus point, we \new{thus} allow users to interactively explore the data and choose their focus point themselves. To support this interaction, we allow users to apply transformations such as rotation and panning. We also allow users to click on the data to select a new focus point, \new{which controls the removal of elements to be rendered at specific scale transitions (as shown in \autoref{tab:hierarchy-mismatch}). First, users can select the focus chromosome (starting at \emph{loci} positions), whose position is the median point within the sequence of fiber positions for that chromosome. This choice controls which chromosome remains as we transition from the \emph{fiber} to the \emph{nucleosome} data scale. Next, starting at the \emph{nucleosome} data scale, users can select a strand of five consecutive \emph{fiber} positions, which then ensures that only this strand remains as we transition from \emph{nucleosome} to \emph{nucleotide} positions.}


To further support the interactive exploration, we also adjust the colors of the elements to be in focus next. For example, the subset of a chromosome next in focus is rendered in a slightly lighter color than the remaining elements of the same level. This approach provides a natural visual indication of the current focus point and guides the view of the users as they explore the scales.

To achieve the \emph{scale-constrained camera navigation}, we measure the distance to a transition or interaction target point in the data sequence. We measure this distance as the span between the camera location and the position of the target level in its currently active scale. This distance then informs the setting of camera parameters and SSAO passes. After the user has selected a new focus point, the current distance to the camera will change, so we adjust also the global scale parameter that we use to control the scale navigation.

\section{Discussion}

Based on our design and implementation we now compare our results with existing visual examples, examine potential application domains, discuss limitations, and suggest several directions for improvement.


\subsection{Comparison to traditionally created illustrations}
\label{sec:comparison}

Measuring the ground truth is only possible to a certain degree, which makes the comparison to ScaleTrotter difficult. One reason is that no static genetic material exists in living cells. Moreover, microscopy is also limited at the scale levels with which we are dealing. We have to rely on the data from the domain experts with its own limitations (\autoref{sec:limitations}) as the input for creating our visualization and compare the results with existing illustrations in both static and animated form.

\begin{figure}[t!]
  \centering
  \centering
	\setlength{\subfigcapskip}{-3ex}%
	\subfigure[\hspace{.535\columnwidth}]{\label{fig:spatial-scale-examples:a}\includegraphics[height=.56\columnwidth]{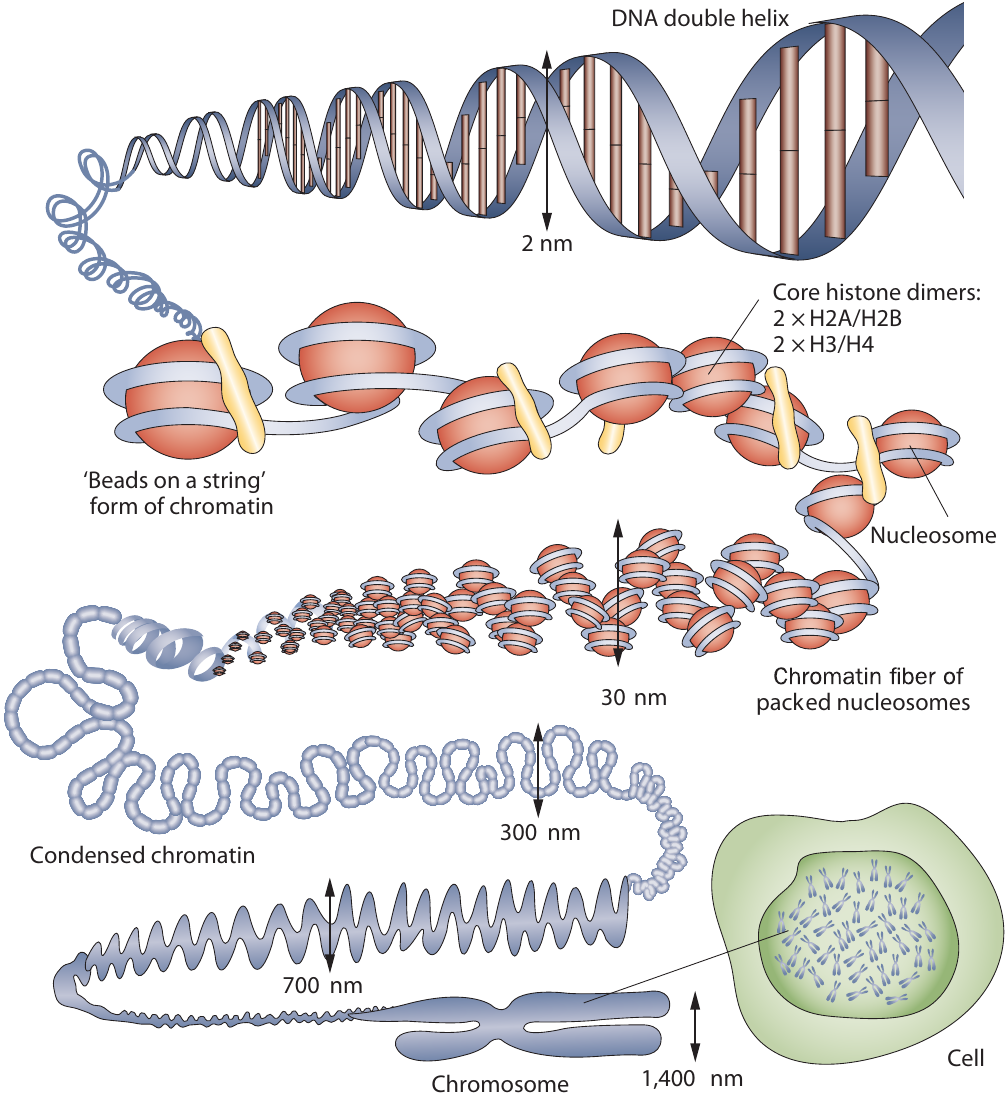}}\hfill%
  \textcolor{white}{\subfigure[\hspace{.37\columnwidth}]{\label{fig:spatial-scale-examples:b}\includegraphics[height=.56\columnwidth]{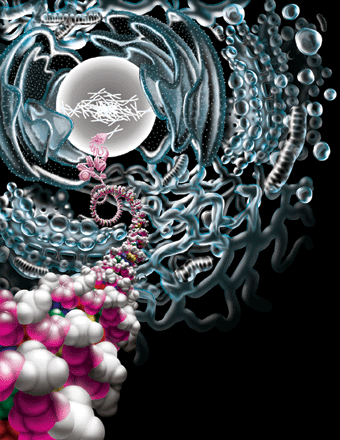}}}\vspace{-1ex}
	\caption{Artistic depictions of image-spatial scale transitions for genome data. \new{Images from \cite{Tonna:2010:MMD}/\hspace{-.2pt}\cite{Pennisi:2001:HG} and \textcopyright\ Springer Nature\discretionary{/}{}{/}The American Association for the Advancement of Science, respectively, used with permission}.}\vspace{-1ex}
	\label{fig:spatial-scale-examples}
\end{figure}

We first look at traditional static multi-scale illustrations as shown in \autoref{fig:spatial-scale-examples}; other illustrations similar to the one in \autoref{fig:spatial-scale-examples:a} can be found in Annunziato's \cite{Annunziato:2008:DPN} and Ou et al.'s \cite{Ou:2017:CV3} works. In \autoref{fig:spatial-scale-examples:a}, the illustrators perform the scale transition along a 1D path, supported by the DNA's extreme length. We do not take this route as we employ the actual positions of elements from the involved datasets. This means that we could also apply our approach to biologic agents such as proteins that do not have an extremely long extent. Moreover, the static illustrations have some continuous scale transitions, \eg, the detail of the DNA molecule itself or the sizes of the nucleosomes. Some transitions in the mul\-ti-scale representation, however, are more sudden such as the transition from the DNA to nucleosomes, the transition from the nucleosomes to the condensed chromatin fiber, and the transition from that fiber to the \SI{700}{\nano\meter} wide chromosome leg. \autoref{fig:spatial-scale-examples:b} has only one such transition. The changeover happens directly between the nucleosome level and the mitotic chromosome. We show transitions between scales interactively using our \emph{visual scale embedding}. The static illustrations in \autoref{fig:spatial-scale-examples} just use the continuous nature of the DNA to evoke the same hierarchical layering of the different scales. The benefit of the spatial scale transitions in the static illustrations is that a single view can depict all scale levels, while our temporally-controlled scale transitions allow us to interactively explore any point in both the genome's spatial layout and in scale. Moreover, we also show the actual physical configuration of every scale according to the datasets that genome researchers provide, representing the current state of knowledge.

We also compare our results to animated illustrations as exemplified by \href{http://www.eamesoffice.com/the-work/powers-of-ten/}{the ``Powers of Ten'' video}\footnote{\href{http://www.eamesoffice.com/the-work/powers-of-ten/}{http://www.eamesoffice.com/the-work/powers-of-ten/} } \cite{Eameses:1977:PTR} and a video treating the composition of the genome\footnote{\href{http://www.wehi.edu.au/wehi-tv/molecular-visualisations-dna}{http://www.wehi.edu.au/wehi-tv/molecular-visualisations-dna} } and created by Drew Berry et al.\ in 2003. The ``Powers of Ten'' video only shows the fibers of the DNA double helix curled into loops---a notion that has since been revised by the domain experts. Nonetheless, the video still shows a continuous transition in scale through blending of aligned representations from the fibers, to the nucleotides, to the atoms. It even suggests that we should continue the scale journey beyond the atoms. The second video, in contrast, shows the scale transitions starting from the DNA double helix and zooming out. The scale transitions are depicted as ``physical'' assembly processes, \eg, going from the double helix to nucleosomes, and from nucleosomes to fibers. Furthermore, shifts of focus or hard cuts are applied as well. The process of assembling an elongated structure through curling up can nicely illustrate the composition of the low-level genome structures, but only if no constraints on the rest of the fibrous structure exist. In our interactive illustration, we have such constraints where we can zoom out and in and where we have restrictions on the locations of all elements coming from the given data. Moreover, the construction also potentially creates a lot of motion due to the dense nature of the genome and, thus, visual noise which might impact the overall visualization. On the other hand, both videos convey the message that no element is static at the small scales. We do not yet show this functionality in our visualizations.

Both static and dynamic traditional visualizations depict the composition of the genome in its mitotic stage. The chromosomes only assume this stage, however, when the cell divides. Our visualization is the first that provides the user with an interactive exploration with smooth scale transitions of the genome in its interphase state, the state in which the chromosomes exist most of the time.


\subsection{Feedback from illustrators and application scenarios}
\label{sec:feedback}

To discuss \new{the creation of illustrations for laypeople with} ScaleTrotter, we asked two professional illustrators for feedback who work on biological and medical visualizations. One of them has ten years experience as a professional scientific illustrator and animator with a focus on biological and medical illustrations for science education. The other expert is a certified illustrator with two years experience plus a PhD in Bioengineering. We conducted a semi-structured interview (approx.\ 60\,min) with them, to get critical feedback \cite{Keefe:2005:ACD,Kosara:2008:VC} on our illustrative multi-scale visualization and to learn how our approach compares to the way they deal with multi-scale depictions in their daily work.


They immediately considered our ScaleTrotter approach for showing genome scale transitions as part of a general story to tell. They missed the necessary additional support for telling a story such as the contextual representation of a cell (for which we could investigate cell\-VIEW \cite{LeMuzic:2015:CEL}) and, in general, audio support and narration. Although they had to judge our results isolated from other story telling methods, they saw the benefits of an interactive tool for creating narratives that goes beyond the possibilities of their manual approaches.

We also got a number of specific pieces of advice for improvement. In particular, they recommended different settings for when to make certain transitions in scale space. The illustrators also suggested the addition of ``contrast'' for those parts that will be in focus next as we zoom in---a feature we then added and describe in \autoref{sec:interaction}.

According to them, our concept of using \emph{visual scale embedding} to transition between different scalar representations has not yet been used in animated illustrations, yet the general concept of showing detail together with context as illustrated in \autoref{fig:grays_anatomy} is known. Instead of using visual scale embedding, they use techniques discussed in \autoref{sec:comparison}, or they employ cut-outs with rectangles or boxes to indicate the transition between scales. Our \emph{visual scale embedding} is seen by them as a clear innovation: ``to have a smooth transition between the scales is really cool.'' Moreover, they were excited about the ability to freely select a point of focus and interactively zoom into the corresponding detail. Basically, they said that our approach would bring them closer to their vision of a ``molecular Maya'' because it is ``essential to have a scientifically correct reference.''
Connected to this point we also discussed the application of ScaleTrotter in genome research. Due to their close collaborations with domain experts they emphasized that the combination of the genomics sequence data plus some type of spatial information will be essential for future research. A combination of our visualization, which is based on the domain's state-of-the-art spatial data, with existing tools could allow genome scientists to better understand the function of genes and certain genetic diseases.

In summary, they are excited about the visual results and see application possibilities both in teaching and in data exploration.

\subsection{\new{Feedback from genome scientists}}
\label{sec:feedbackexpert}

As a result of our conversation with the illustrators they also connected us to a neurobiologist who investigates 3D genome structures at single cell levels, \eg, by comparing cancerous with healthy cells. His group is interested in interactions between different regions of the genome. Although the spatial characteristics of the data are of key importance to them, they still use 2D tools. The scientist confirmed that a combination of their 2D representations with our interactive 3D-spatial multi-scale method would considerably help them to understand the interaction of sequentially distant but spatially close parts of the genome, processes such as gene expression, and DNA-protein interactions.

\new{We also presented our approach to an expert in molecular biology with 52 years of age and 22 years of post-PhD experience. He specializes in genetics and studies the composition, architecture, and function of SMC complexes. We conducted a se\-mi-struc\-tured interview (approx.\ 60 minutes) to discuss our results. He stated that transitions between several scales are definitely useful for analyzing the 3D genome. He was satisfied with the coarser chromosomes and loci representations, but had suggestions for improving the nucleosome and atomic scales. In particular, he noted the lack of proteins such as histones. He compared our visualization with existing electron microscopy images \cite{Ou:2017:CV3,Olins:2003:CHV}, and suggested that a more familiar \textit{fi\-la\-ment-like} representation could increase understandability. In his opinion, some scale transitions happened too early (\eg, the transition from chro\-mo\-some-co\-lored to nucleotide-colored nucleotides). We adjusted our parametrization accordingly. In addition, based on his feedback, we added an interactive \emph{scale offset} control that now allows users to adjust the scale representation for a given zoom level. This offset only adapts the chosen representation according to \autoref{tab:hierarchy-mismatch}, while leaving the size on the screen unchanged. The expert also suggested to build on the current approach and extend it with more scales, which we plan to do in the future. Similar to the feedback from the neurobiologist, also the molecular biologist agrees that an integration with existing 2D examination tools has a great potential to improve the workflow in a future visualization system. }

\subsection{Limitations}
\label{sec:limitations}

There are several limitations of our work, the first set relating to the source data. While we used actual data generated by domain experts based on the latest understanding of the genome, it is largely generated using simulations and not actual measurements (\autoref{sec:implementation:data}. We do not use actual sequence data at the lowest scales. Moreover, our specific dataset only contains 45 chromosomes, instead of the correct number of 46. We also noticed that the dataset contains 23,958,240 nucleosome positions, yet when we multiply this with the sum of 146 base pairs per nucleosome we arrive at $\approx 3.5$\,Gb for the entire genome---not even including the linker base pairs in this calculation and for only 45 chromosomes. Ultimately better data is required. The overall nature of the visualization and the scale transitions would not be impacted by the modified data and we believe that the data quality is already sufficient for general illustration and teaching purposes.

Another limitation is the huge size of the data. Loading all positions for the interactive visualization takes approx.\ \new{two minutes, but we have not yet explored the feasibility of also loading actual sequence data. We could investigate loading} data on-demand for future interactive applications, in particular in the context of tools for domain experts. For such applications we would also likely have to reconsider our design decision to leave out data in the detailed scales, as these may interact with the parts that we do show. We would need to develop a space-dependent look-up to identify parts from the entire genome that potentially interact with the presently shown focus sequences.
Another limitation relates to the selection of detail to zoom into. At the moment, we determine the focus interactively based on the currently depicted scale level. This makes it, for example, difficult to select a chromosome deep inside the nucleus or fibers deep inside a chromosome. A combination with an abstract data representation---for example with a domain expert sequencing tool---would address this problem.


\subsection{Future work}
\label{sec:future}
Beyond addressing the mentioned issues, we would like to pursue a number of additional ideas in the future. A next step towards adoption of our approach in biological or medical research is to build an analytical system on top of ScaleTrotter that allows us to query various scientifically relevant aspects. As noted in \autoref{sec:feedback}, one scenario are spatial queries to determine whether two genes are located in a close spatial vicinity in case they somehow are related. Other visualization systems developed in the past for analyzing gene expressions can benefit from the structural features that ScaleTrotter offers.


Extending to other subject matters, we will also have to investigate scale transitions where the scales cannot be represented with sequences of blobs. For example, can we also use linear or volumetric representations and extend our visual space embedding to such structures? Alternatively, can we find more effective scale transitions to use such as geometry-based ones (\eg, \cite{Miao:2018:MVS, Zwan:2011:IMV, Lueks:2011:SCC}), in addition to the visual embedding and the color changes we use so far? We have to avoid over-using the visual variable color which is a scarce resource. Many elements could use color at different scales, so dynamic methods for color management will be essential.

Another direction for future research are generative methods for completing the basic skeletal genetic information on the fly. Currently we use data that are based on positions of nucleotides, while higher-level structures are constructed from these. Information about nucleotide orientations and their connectivity is missing, as well as the specific sequence which is currently not derived from real data. ScaleTrotter does not contain higher-level structures and protein complexes that hold the genome together and which would need to be modeled with a strict scientific accuracy in mind. An algorithmic generation of such models from Hi-C data would allow biologists to adjust the model parameters according to their mental model, and would give them a system for generating new hypotheses. Such a generative approach would also integrate well with the task of adding processes in which involve the DNA, such as condensation, replication, and cell division.

A related fundamental question is how to visualize the dynamic characteristics of the molecular world. It would be highly useful to portray the transition between the interphase and the mitotic form of the DNA, to support visualizing the dynamic processes of reading out the DNA, and to even show the Brownian motion of the atoms.  

Finally, our visualization relies on dedicated decisions of how to parameterize the scale transitions. While we used our best judgment to adjust the settings, the resulting parameterization may not be universally valid. An interactive illustration for teaching may need parameters different from those in a tool for domain experts. It would be helpful to derive templates that could be used in different application contexts.

\section{Conclusion}

ScaleTrotter constitutes one step towards understanding the mysteries of human genetics---not only for a small group of scientists, but also for larger audiences. It is driven by our desire as humans to understand \href{https://gutenberg.spiegel.de/buch/faust-eine-tragodie-3664/4}{\emph{``was die Welt im Innersten zusammenh{\"a}lt''}} [what \href{https://www.gutenberg.org/files/14591/14591-h/14591-h.htm#I}{``binds the world, and guides its course''}] \cite{Goethe:1808:FTE}.
We believe that our visualization has the potential to serve as the basis of teaching material about the genome and part of the inner workings of biologic processes. It is intended both for the general public and as a foundation for future visual data exploration for genome researchers. In both cases we support, for the first time, an interactive and seamless exploration of the full range of scales---from the nucleus to the atoms of the DNA.

From our discussion it became clear that such multi-scale visualizations need to be created in a fundamentally different way as compared to those excellent examples used in the astronomy domain. In this paper we thus distinguish between the \emph{positive-exponent scale-space} of astronomy (looking inside-out) and the negative-exponent scale-space of genome data (looking outside-in). For the latter we provide a multi-scale visualization approach based on \emph{visual scale embedding}.
We also discuss an example on how the controlled use of abstraction in (illustrative) visualization allows us to employ a space-effi\-cient superimposition of visual representations. This is opposed to juxtaposed views \cite{Viola:2018:PCA}, which are ubiquitous in visualization today.

A remaining question is whether the tipping point between the different types of scale spaces is really approximately one meter \mbox{($1 \cdot 10^0$\,m)} or whether we should use a different point in scale space such as \SI{1}{\milli\metre}. The answer to this question requires further studies on how to illustrate multi-scale subject matter. An example is to generalize our approach to other biologic phenomena such as mitotic DNA or microtubules as suggested in \autoref{sec:future}. If we continue our journey down the negative-exponent scale-space we may discover a third scale-space region. Models of atoms and subatomic particles seem to again comprise much empty space, similar to the situation in the positive-exponent scale-space. A bigger vision of this work thus is to completely replicate \href{http://www.eamesoffice.com/the-work/powers-of-ten/}{the ``Powers of Ten'' video}---the 36 orders of magnitude from the size of the observable universe to sub-atomic particles---but with an interactive tool and based on current data and visualizations.


\acknowledgments{
\new{We thank the genome scientists who provided the data that our tool relies on and who answered our questions about it. We also thank everyone who provided feedback about our approach. Part of this work was funded under the ILLUSTRARE grant by both the Austrian Science Fund (FWF): I~2953-N31 and the French National Research Agency (ANR): ANR-16-CE91-0011-01. The research was further supported by funding from King Abdullah University of Science and Technology (KAUST), under award number BAS/1/1680-01-01 and by funding from ILLVISATION grant by WWTF (VRG11-010). Authors would like to thank Nanographics GmbH (\href{http://nanographics.at/}{nanographics.at}) for providing the Marion Software Framework. This paper was partly written in collaboration with the VRVis Competence Center. VRVis is funded by BMVIT, BMWFW, Styria, SFG and Vienna Business Agency in the scope of COMET -- Competence Centers for Excellent Technologies (854174), which is managed by FFG.}}

\bibliographystyle{abbrv-doi-hyperref-narrow}

\bibliography{template}

%
%

\end{document}